\documentclass[iicol,sn-mathphys-num]{sn-jnl}
\usepackage[utf8]{inputenc}
\usepackage{subcaption}
\usepackage{threeparttable}
\usepackage{textgreek}
\usepackage{booktabs}
\usepackage{csquotes} 
\usepackage[version=4]{mhchem}
\newcommand{\cawo}{\ce{CaWO_4}}
\usepackage[separate-uncertainty=true, free-standing-units=true]{siunitx}
\DeclareSIUnit\eVc{\eV\per\clight\squared}
\DeclareSIUnit\clight{\text{\!\ensuremath{c}}}
\usepackage{cleveref}
\newcommand{\geant}{\textsc{Geant4}}
\newcommand{\cresst}{\textsc{Cresst}}

\title{Impact of \geant{}'s Electromagnetic Physics Constructors on Accuracy and Performance of Simulations for Rare Event Searches}

\begin{document}
\author*[1,2]{\fnm{H.} \sur{Kluck}}\email{holger.kluck@oeaw.ac.at}
\author[3]{\fnm{R.} \sur{Breier}}
\author[1,2]{\fnm{A.} \sur{Fuß}}
\author[1]{\fnm{V.} \sur{Mokina}}
\author[3,4,5]{\fnm{V.} \sur{Palušov\'a}}
\author[3]{\fnm{P.} \sur{Povinec}}
\affil*[1]{\orgname{Institut f\"ur Hochenergiephysik der \"Osterreichischen Akademie der Wissenschaften}, \orgaddress{\city{Wien}, \postcode{1050}, \state{Austria}}}
\affil[2]{\orgdiv{Atominstitut}, \orgname{Technische Universit\"at Wien}, \orgaddress{\city{Wien}, \postcode{1020}, \state{Austria}}}
\affil[3]{\orgdiv{Faculty of Mathematics, Physics and Informatics}, \orgname{Comenius University}, \orgaddress{\city{Bratislava}, \postcode{84248}, \state{Slovakia}}}
\affil[4]{\orgdiv{Institute of Experimental and Applied Physics}, \orgname{Czech Technical University in Prague}, \orgaddress{\city{Prague 1}, \postcode{110 00}, \state{Czech Republic}}}

\affil[5]{Present address: \orgdiv{Institut f\"ur Physik}, \orgname{Johannes Gutenberg-Universit\"at Mainz}, \orgaddress{\city{Mainz}, \postcode{55128}, \state{Germany}}}

\abstract{A primary objective in contemporary low background physics is the search for rare and novel phenomena beyond the Standard Model of particle physics, e.g. the scattering off of a potential Dark Matter particle or the neutrinoless double beta decay. The success of such searches depends on a reliable background prediction via Monte Carlo simulations. A widely used toolkit to construct these simulations is \geant{},
which offers the user a wide choice of how to implement the physics of particle interactions. For example, for electromagnetic interactions, \geant{} provides pre-defined sets of implementations: physics constructors.
As decay products of radioactive contaminants contribute to the background mainly via electromagnetic interactions, the physics constructor 
used in a \geant{} simulation may have an impact on the total energy deposition inside the detector target. To facilitate the selection of physics constructors for simulations of experiments that are using
\cawo{} and Ge targets, we quantify their impact on the total energy deposition for several test cases.
These cases consist of radioactive contaminants commonly encountered, covering energy depositions via \textalpha{}, \textbeta{}, and \textgamma{} particles, as well as two examples for the target thickness: thin and bulky. We also consider the computing performance of the studied physics constructors.}


\maketitle

\section{\label{sec:introduction}Introduction}

\cawo{} and Ge are common target materials for a wide range of experiments that are searching for rare and novel phenomena: For example the scattering off of a potential Dark Matter particle (e.g.\ \textsc{Cresst} \cite{CRESST:2019jnq} for \cawo{} or CDMSlite \cite{Alkhatib2021} for Ge), the Coherent Elastic Neutrino-Nucleus Scattering (CE\textnu{}NS) (e.g.\ \textsc{Nucleus} \cite{Angloher2019} for \cawo{} or \textsc{Conus} \cite{Ackermann2024} for Ge), or the neutrinoless double beta (0\textnu{}2\textbeta{}) decay (e.g.\ the future \textsc{Legend} experiment \cite{Abgrall2017} for Ge).

By their very nature, i.e., searching for \emph{rare} events amidst a usually more frequent background, such kinds of experiment depend crucially on a reliable and verified background prediction. Typically, these predictions are based on Monte Carlo simulations of relevant background sources. A widely used toolkit to create these simulations is \geant{} \cite{GEANT4:2002zbu,Allison:2006ve,Allison:2016lfl}, which allows its user to consider a given physics interaction by implementing this interaction as a \emph{physics process} that relies on one or more \emph{physics models} in a so-called \emph{physics list}. For electromagnetic interactions (EM), e.g.\ ionisation, photo absorption, Compton scattering, etc., \geant{} provides several electromagnetic \emph{physics constructors}: they are predefined sets of processes, models, and their parameters settings to enable a consistent interplay between EM interactions, see the \geant{} manual \cite{Geant4Manual} for details. Hence, a physics list may include directly a set of physics processes and models or include physics constructors, which include physics processes and models, or any combinations thereof.

In the literature, validation studies of the predefined physics lists and physics constructors of \geant{} are reported for a wide range of physics interactions and observables, e.g. atomic relaxation \cite{Guatelli2007}, electron energy deposition \cite{Lechner2009, Seo2011, Batic2013}, electron back scattering \cite{Batic2013,Basaglia2015,Basaglia2016a} and electromagnetic photon interaction \cite{Cirrone2010}. Also for the medical physics user community dedicated studies exists, e.g.\ \cite{Arce:2020}. However, to our knowledge, no such study exists on the impact of the physics constructor or physics list used on the total energy deposited by radioactive decays in \cawo{} or Ge targets. As pointed out in \cite{Basaglia2015a, Basaglia2016a}, an observable such as total energy deposition is the result of several physics interactions (e.g. atomic relaxation, photon interaction, etc.). The relevance of each of these interactions and the precision of the related implementations depends on the actual involved materials, energies, and geometries, i.e. on the actual use case. Usually, it is precarious to extrapolate the observable from studies based on one use case (e.g. monochromatic X-ray transmission through thin targets) to another use (e.g. energy deposition by \textgamma{} rays caused by radioactive contaminations in thick targets). A physics constructor that provides a precise simulation for the first use case may provide a less precise simulation for the last use case.

Based on earlier, preliminary work \cite{IDM2022}, we provide a dedicated study for the use case of total energy deposit by radioactive contaminations. As the products of the radioactive decays deposit their energy mainly via electromagnetic interactions, we examine the impact of different electromagnetic physics constructors on the total energy deposition for our test cases, i.e.\ combinations of radioactive contaminants, target material (\cawo{} and Ge) and target thickness. The construction and study of new dedicated physics lists or physics constructors is beyond the scope of this work and maybe the topic of future investigations.
The aim of this work is to give an assessment to what extent the selected physics constructor affects the simulated observable of total energy deposition, i.e. how compatible different physics constructors are\footnote{Following the terminology as defined in \cite{Guatelli2007}, we note that this work is neither a \emph{verification}, i.e. an assessment of the accuracy of an implementation, nor a \emph{validation}, i.e. a comparison of an implementation with experimental reference data.}. This gives a notion of the systematic uncertainty of the simulation caused by the possibility to chose different physics constructors. To assess the compatibility in a qualitative and objective way, we adopt the methodology established in \cite{Guatelli2007,Lechner2009,Batic2013,Basaglia2016a}. To give some guidance which physics constructor to choose in case of compatibility, we consider also the computing performance. If two physics constructors are compatible, one may choose the one that requires less computation resources.

In \cref{sec:configurations} we motivate our test cases and describe their implementation before we list the studied physics constructors and their settings in \cref{sec:physicslists}. In \cref{sec:statistic} we outline our concrete implementation of the adopted methodology to judge the statistical consistency. We report the outcome of the tests in terms of statistical compatibility in \cref{sec:results} and in terms of computing performance in \cref{sec:performance}. Finally, we conclude in \cref{sec:conclusion}.

\section{\label{sec:configurations}Test Cases }

The relevant test cases are defined by three characteristics: the target material, its geometry, and the radionuclide that contaminate the target.

Based on the reasoning in \cref{sec:introduction}, we decided on \cawo{} and \ce{Ge} as target materials.

As radionuclides, we select six common contaminants: 
low $Q$-value \textbeta{} emitters (\ce{^{228}Ra} and \ce{^{210}Pb} with $Q = \mathcal{O}(\qty{10}{\keV})$), high $Q$-value \textbeta{} emitters (\ce{^{208}Tl} and \ce{^{210}Tl} with $Q = \mathcal{O}(\qty{1}{\MeV})$) and
\textalpha{} emitters (\ce{^{211}Bi} and \ce{^{234}U}). Five nuclides feature \textgamma{} lines with intensities $I_\gamma > \qty{1}{\percent}$ that cover the energy range from $\mathcal{O}(\qty{10}{\keV})$ to $\mathcal{O}(\qty{1}{\MeV})$, see \cref{tab:contaminants} for their properties. With this selection we cover electron, gamma and ion interactions, i.e. the most relevant interactions for electromagnetic background from natural radioactivity. We note that the neutronic background is outside the scope of this work.

\begin{table}[ht]
    \caption{\label{tab:contaminants}Decay properties of the simulated nuclides: dominant decay mode and associated $Q$-value, and the energy $E_\gamma$ and absolute intensity $I_\gamma$ of the most prominent \textgamma{} line of the decay (for $I_\gamma > \qty{1}{\percent}$). Data taken from \cite{LiveChartOfNucllide}.}
    \centering
    \begin{tabular}{c c S S S}\toprule
        Nuclide & Decay & {$Q$ / \unit{\keV}} & {$E_\gamma$ / \unit{\keV}} & {$I_\gamma$ / \unit{\percent}}\\ \midrule
        \ce{^{228}Ra} & \textbeta{}  & 45.5   & 13.52    &  1.6 \\
        \ce{^{210}Pb} & \textbeta{}  & 63.5   & 46.539   & 4.25 \\
        \ce{^{234}U}  & \textalpha{} & 4857.5 & {--}     & {--} \\
        \ce{^{208}Tl} & \textbeta{}  & 4998.4 & 2614.511 & 99.754 \\
        \ce{^{210}Tl} & \textbeta{}  & 5481   & 799.6    & 98.96 \\
        \ce{^{211}Bi} & \textalpha{} & 6750.4 & 351.07   & 13.02 \\ \bottomrule
    \end{tabular}
\end{table}

\begin{figure*}[ht]
\centering
\subfloat[][]{\includegraphics[width=0.45\textwidth]{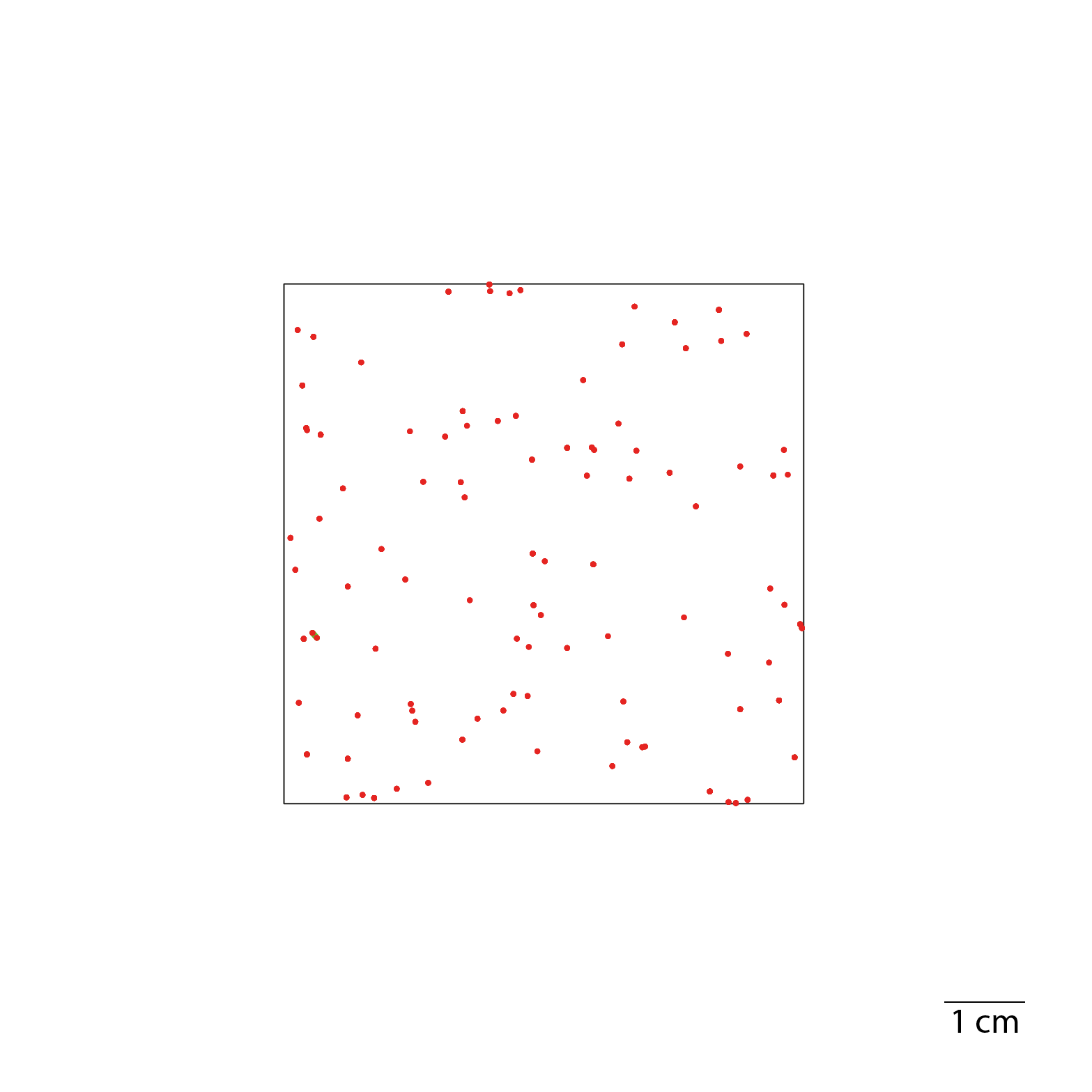}\label{fig:geometry:bulky}}~
\subfloat[][]{\includegraphics[width=0.45\textwidth]{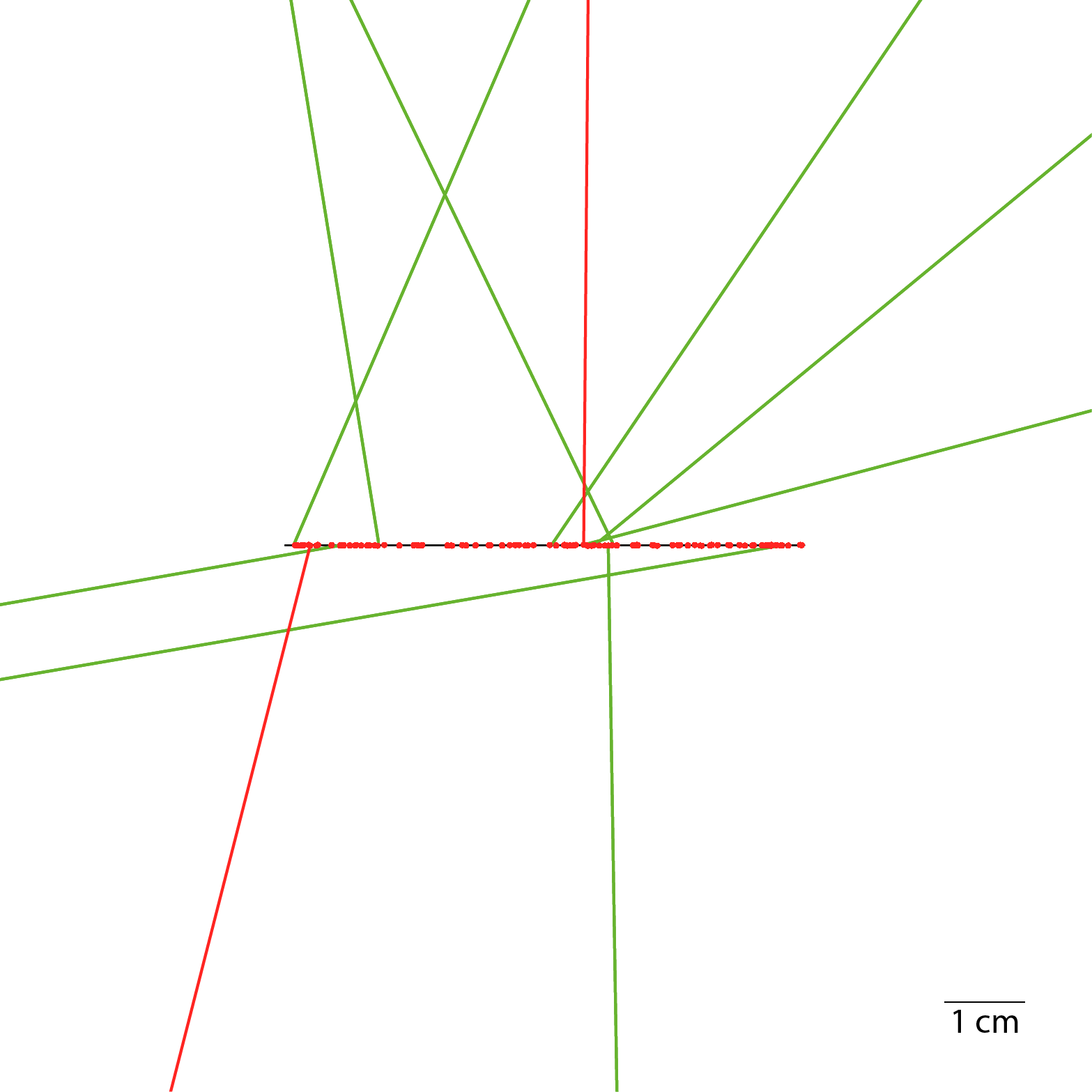}\label{fig:geometry:thin}}
\caption{\label{fig:geometry}\geant{} visualisation of \num{100} \textbeta{} decays of \ce{^{210}Pb} ($Q$-value of \qty{63.5}{\keV}) each in \cawo{} targets with a quadratic cross section of \qtyproduct{64 x 64}{\mm} in the $x$-$y$-plane and for two thicknesses along the $z$-axis: (\protect\subref{fig:geometry:bulky}) \qty{64}{\mm} for the \emph{bulky} target and (\protect\subref{fig:geometry:thin}) \qty{100}{\um} for the \emph{thin} target; the targets are placed in vacuum. Gamma rays are shown in \emph{green} and electron tracks in \emph{red}. In case of the bulky target, all decay products are absorbed inside the target; in case of the thin target, some low energy decay products can leave the target. If the decay products get immediate absorbed, their tracks are reduced to \emph{dots}. Due to technicalities of the visualisation, green dots, representing gamma rays, are usually covered by the red dots, representing electrons, and hence mostly not visible.}
\end{figure*}

As geometry, we chose a cuboid with a cross-section of \qtyproduct{64 x 64}{\mm} in two configurations: \textit{bulky} with a thickness of \qty{64}{\mm} and \textit{thin} with a thickness of \qty{100}{\um}, see \cref{fig:geometry}. The two configurations represent the two extreme cases that can occur in rare event searches, motivated by the largest and smallest distinct part of a \cresst{} detector module, i.e. the absorber crystal and a bond wire, respectively. Although the target thickness is an experiment-specific value and hence the adopted values are to some degree arbitrary, it allows us to study the impact of the target thickness on the local energy absorption. As demonstrated in \cref{fig:geometry:bulky}, the products from low-$Q$ decays are efficiently absorbed by the bulky target, while they leak out from the thin target in \cref{fig:geometry:thin}. Therefore, the latter case is more sensitive to the energy deposited by individual particle interactions inside the target, whereas the first case is more sensitive to the total energy released in the nuclear decay, i.e.\ the $Q$-value.

\section{\label{sec:physicslists}Studied Electromagnetic Physics Constructors}
\geant{} provides several predefined sets of EM physics processes, models, and their parameter settings called EM physics constructors, which offer users an easy way to consistently implement a set of EM interactions. Often they are tuned for specific use cases, e.g.\ higher simulation performance but less low energy precision (\texttt{G4EmStandardPhysics\_option1}) or vice versa (\texttt{G4EmStandardPhysics\_option4}). A further advantage of using these predefined physics constructors is that they are tested on a regular basis by the \geant{} developers. The composition of a physics constructor, i.e.\ what physics processes and models are included, and how their parameters are set, can change between different \geant{} versions. In the following, we use \geant{} version 10.6.3.

Twelve physics constructors were selected to be used in this work, see~\cref{table:constructor_comparison} for details.
As it has a large impact on computing performance (see \cref{sec:performance}), we note that there are three approaches to implement \emph{Coulomb scattering} of charged particles \cite{Ivanchenko2010}:

(i) A \emph{single-scattering} approach, where each Coulomb scattering is explicitly simulated. This is computationally expensive, as it results in numerous small simulation steps. \geant{} offers two physics models for single scattering: \texttt{G4eCoulombScatteringModel} (based on \cite{FernandezVarea1993,Wentzel1926,Bethe1953}) and \texttt{G4eSingleCoulombScatteringModel} \cite{Boschini2013}. The \texttt{G4EmStandardSS} constructor uses exclusively the latter model.

(ii) A \emph{multiple-scattering} (MSC) approach: based on a multiple-scattering theory, the net displacement, energy loss, and change of direction of the incident particle after several scatterings are calculated in one step \cite{Urban2006}. Compared to the single-scattering approach, this approach is less computationally expensive, as it reduces the number of simulation steps. However, the correct treatment of large-angle scattering is difficult. Furthermore, the MSC models have several parameters whose settings may have a strong impact on their accuracy and performance. \geant{} offers two physics models for multiple scattering of electrons and positrons: \texttt{G4UrbanMscModel} (based on \cite{Lewis1950,Urban2006}) and \texttt{G4GoudsmitSaundersonMscModel} \cite{Novak:2025}. The former is used by \texttt{G4EmStandardPhysics}, \texttt{G4EmStandardPhysics\_option1}, \texttt{G4EmStandardPhysics\_option2}, and \texttt{G4EmStandardPhysics\_option3}; the latter is used by \texttt{G4EmStandardPhysics\_option4}, \texttt{G4EmPenelopePhysics}, \texttt{G4EmLivermorePhysics}, and \texttt{G4EmStandardPhysicsGS}. Except for \texttt{G4EmStandardPhysics\_option3}, all mentioned constructors use additional models at higher energies: \texttt{G4eSingleCoulombScatteringModel} and \texttt{G4WentzelVIModel} (see below). \texttt{G4EmStandardPhysics\_option4}, \texttt{G4EmPenelopePhysics}, and \texttt{G4EmLivermorePhysics} use the same parameter settings for the \texttt{G4GoudsmitSaundersonMscModel}.

(iii) A hybrid approach, where single scattering is used for the treatment of large-angle scattering and multiple scattering otherwise. With decreasing kinetic energy of the incident particle, the relative contribution of large-angle scattering increases. Hence, the performance of this approach approximates that of the single-scattering approach. \geant{} offers the \texttt{G4WentzelVIModel} and related models (based on \cite{Wentzel1926,FernandezVarea1993,Urban2006,Lewis1950}). The \texttt{G4EmStandardPhysicsWVI} and \texttt{G4EmLowEPPhysics} constructors use exclusively this model.

We refer to the \geant{} physics manual \cite{Geant4PhysicsManual} for a more in-depth discussion of the internal structure of the physics constructors, and their similarities and differences.
 
In \geant{}, the simulated physics is not only determined by the chosen physics models but also by the so-called \emph{production cut} or range cut value: potential secondary particles that have too little kinetic energy to travel further than the production cut value are not produced as a distinct particle track, but their kinetic energy is deposited locally \cite[pp.~17, 208, 255]{Geant4Manual}. Usually, this cut is applied only to ionisation and bremsstrahlung, however \texttt{G4EmStandardPhysics\_option1} applies this cut to all EM interactions. Lower values of the production cut increase the precision of particle transport, e.g.\ when secondary decay products may \textit{leak} out of \textit{thin} targets. Hence, usually the production cut should be smaller than the dimension of the target.  However, the drawback of a smaller production cut is the slower simulation performance.

In order to study the impact of the production cut value on our test cases and on the performance of the simulation, each EM constructor was applied with five production cut values:\footnote{The actual values are somewhat arbitrary and were motivated by typical length scales of CRESST detector components, from small to large dimensions: bonding wire, thickness of reflecting foil, TES, absorber crystal, and detector module.} \qty{100}{\nm}, \qty{1}{\um}, \qty{1}{\mm}, \qty{1}{\cm}, \qty{10}{\cm}. In this work, the used physics constructor $P_i$ and the applied cut value $c_j$ characterised one \geant{} physics \textit{configuration} $\pi_{ij}=(P_i,c_j)$, yielding together 60 physics configurations (=12 physics constructors $\times$ 5 production cut values).

As \texttt{G4EmStandardPhysics$\_$option4} is regarded as the most accurate EM constructor\footnote{See \cite[p.~215]{Geant4Manual}: \textquote{\emph{G4EmStandardPhysics\_option4}, containing the most accurate models from the Standard and Low Energy Electromagnetic physics working groups.}} we chose it together with \geant{}'s default value for the production cut of \qty{1}{\mm} \cite[p.~17]{Geant4Manual} as our \emph{reference configuration} $\pi_\mathrm{ref}=(P_\mathrm{ref},c_\mathrm{ref})$.
\begin{table*}[ht]
\caption{List of default and optional EM physics constructors selected for this work with comments. For more details of each of them, please read~\cite{Allison:2016lfl, Apostolakis:2009egq, Dotti:2011zz, Geant4PhysicsManual,Arce:2020}.}
{
\centering
\begin{tabular*}{\textwidth}{@{\extracolsep{\fill}} p{0.3\textwidth} p{0.65\textwidth}}
\toprule
\textbf{Physics constructor}        & \textbf{Comment}                           \\
\midrule
\texttt{G4EmStandardPhysics}          & Default in \geant{}, used by ATLAS    \\
\hline
\texttt{G4EmStandardPhysics\_option1} & High energy physics, simplified, fast but not precise for sampling calorimeters, used by CMS   \\
\hline
\texttt{G4EmStandardPhysics\_option2} & Experimental, high energy physics, simplified, fast but not precise for sampling calorimeters, used by LHCb   \\
\hline
\texttt{G4EmStandardPhysics\_option3} & For space science applications, detailed, standard models                \\
\hline
\texttt{G4EmStandardPhysics\_option4}\footnotemark[1] & Optimal mixture for precision, most accurate EM models and settings, best combination, for medical application \cite{Arce:2020} \\
\hline
\texttt{G4EmLivermorePhysics}       & Detailed, based on \texttt{G4EmStandardPhysics\_option4} constructor, using Livermore models which describe the interactions of electrons and photons with matter down to about \qty{250}{\eV} (close to the K-shell Auger peak from C) using interpolated data tables based on the Livermore library.  \\
\hline
\texttt{G4Em\-Livermore\-Polarized\-Physics} & The same as \texttt{G4EmLivermorePhysics} using polarized models. \\
\hline
\texttt{G4EmPenelopePhysics}          & Based on \texttt{G4EmStandardPhysics\_option4} constructor, using specific low-energy Penelope models below \qty{1}{\GeV}.  \\
\hline
\texttt{G4EmLowEPPhysics}             & This configuration is the same as in the default \texttt{G4EmStandardPhysics} constructor, with using new low energy models. \\
\hline
\texttt{G4EmStandardPhysicsWVI}              & This configuration is the same as in the default \texttt{G4EmStandardPhysics} constructor, except \texttt{G4WentzelVIModel} is used for simulation of multiple scattering combined with single elastic at large angles. For ion ionisation of ions below 2 MeV/u the Bragg model is used, above 2 MeV/u the ATIMA model is applied (\texttt{G4AtimaEnergyLossModel}) with ATIMA fluctuation model (\texttt{G4AtimaFluctuations}). \\ 
\hline
\texttt{G4EmStandardSS}\footnotemark[2]             & Configuration is the same as in the default \texttt{G4EmStandardPhysics} constructor, except multiple scattering is not used, and only elastic scattering process is applied for all changed particles.  \\
\hline
\texttt{G4EmStandardPhysicsGS}               & This configuration is the same as in the default \texttt{G4EmStandardPhysics} constructor, except multiple scattering of e-
 and e+, which is handled by the Goudsmit-Sounderson model from \qtyrange{0}{100}{\MeV}.  \\
\bottomrule
\end{tabular*}%
}
\par{
\textsuperscript{1}In this work, we use this physics constructor as a reference to which the other physics constructors are compared.\par
\textsuperscript{2}Albeit the name of this physics constructor itself is given as \emph{G4EmStandardSS}, the corresponding source code files have the name \emph{G4EmStandardPhysicsSS}.
}   
\label{table:constructor_comparison}
\end{table*}

In addition, for each configuration we simulated each of the 24 test cases (= 2 targets $\times$  2 thicknesses $\times$ 6 contaminants), yielding together 1440 data sets (= 24 test cases $\times$ 60 configurations).

\section{\label{sec:statistic}Determining Statistical Compatibility}

To determine the compatibility of the reference physics configuration $\pi_\mathrm{ref}$ with the remaining other 59 physics configurations in terms of total energy deposition inside the target, we compare the simulated spectra for all 24 test cases. To quantify the compatibility, we apply two stages of statistical tests:
In the first stage (see \cref{sec:statistic:gof}), we address the question of how well the spectrum obtained with the reference physics configuration $\pi_\mathrm{ref}$ can be described by the spectrum of any other physics configuration $\pi_{ij}$ 
by appropriate \emph{goodness-of-fit} (GoF) tests. The second stage (see \cref{sec:statistic:cat}) is based on categorical tests, which determine if the difference in compatibility observed across our physics configurations $\pi_{ij}$ is statistically significant. 
As the statistical tests are not uniformly sensitive to differences in the spectra at all energies, a variety of statistical tests is applied in each step of the analysis to minimise the possibility of introducing systematic effects.

\subsection{Compatibility With Reference Physics Configuration}\label{sec:statistic:gof} 
We use GoF tests to determine the statistical compatibility between the spectra simulated by the reference physics configuration $\pi_\mathrm{ref}$ and any other physics configuration $\pi_{ij}$; hereafter we call them \emph{reference spectrum} and \emph{test spectra}. The initial null hypothesis $H_0$ is that both data sets follow the same distribution. In a GoF test, to accept or reject the null hypothesis, we construct a test statistic whose value is sensitive to the measure of agreement between the data set and the predictions of our hypothesis $H_0$.

The $p$-value of a GoF test is a probability of obtaining data that are as compatible with our prediction as the reference data, or less. The significance level $\alpha$ was chosen at \qty{5}{\percent}, a conventional value based on Fisher's argument that a chance of 1 in 20 represents an unusual sampling occurrence \cite{kimjae2015}, and the hypothesis that two histograms follow an identical distribution is rejected if the $p$-value is lower than $\alpha$. 

Three independent GoF tests were performed to test this compatibility: Kolmogorov-Smirnov (KS) \cite{Kolmogorov1933, Smirnov1939}, $\chi^2$ test \cite{Pearson1900, Cochran1952}, and Anderson-Darling (AD) \cite{Anderson1956}. As these tests use different measures of discrepancy between the reference and test data, each one has its own advantages and disadvantages: KS test suffers from a sensitivity loss at the boundaries, because it tends to be more sensitive near the centre of the spectrum than at the tails. Thus, if the discrepancies between the reference and test data are primarily in the tails of the spectrum, the $\chi^2$ test may perform better. The disadvantage of the $\chi^2$ test is that the value of its test statistic depends on how the data is binned. The AD test gives more weight to the tails and thus is more sensitive to deviations in the tails of the spectrum. As pointed out in \cite{Kim2015,Basaglia2015a}, relying on only one GoF test can result in a systematic bias, hence we will state the results of all three GoF tests in the following.

The outcome of the GoF tests is reported in the form of an \emph{efficiency} $\xi_{ij}=n_{ij,\mathrm{acc}}/(n_{ij,\mathrm{acc}}+n_{ij,\mathrm{rej}})$ that is defined as the fraction of test cases where $\mathrm{H_0}$ is \emph{accepted}, i.e.\ the $p$-value resulting from the GoF tests is larger than the significance level $\alpha$. This quantifies the capability of a physics configuration $\pi_{ij}$ to produce results which are statistically consistent with the simulated outcome of our reference physics configuration $\pi_\mathrm{ref}$. Consequently, $1-\xi_{ij}$ is the fraction of test cases where $H_0$ is \emph{rejected}. Those are the cases where a physics configuration $\pi_{ij}$ produces results which are statistically inconsistent with the one of our reference physics configuration $\pi_\mathrm{ref}$. The uncertainty of a given efficiency was calculated based on Bayes' theorem, following the description in \cite{Paterno:2004cb}. This method is more appropriate than conventional binomial errors, which yields the absurd result of vanishing uncertainty in the limiting cases when efficiency is either \qty{100}{\percent} or \qty{0}{\percent}.

\subsection{Compatibility Between Physics Configurations}\label{sec:statistic:cat}
Categorical statistical tests summarise the relationship between the results of the GoF tests. These results are categorised into 2 outcomes: either the hypothesis of compatibility was \emph{rejected} or \emph{accepted}.
Categorical variables represent counts or frequencies, like in our case the efficiencies $\xi_{ij}$, that may be based on \emph{paired} (dependent) or \emph{unpaired} (independent) samples. Simulations that use different physics constructors generate unpaired samples, while simulations that use the same physics constructor but vary in a secondary feature, here: the production cut, produce paired samples\footnote{We treat each physics constructor as its own one parametric-model with the production cut for secondary particle production as a free parameter.}.

For convenience, frequencies as categorical variables are arranged in contingency tables and their layout is different for paired or unpaired samples. 
For paired data, the appropriate $2\times 2$ contingency tables are built by counting the number of cases where for both physics configurations $\pi_{ij}$, $\pi_{ik}$ the compatibility hypothesis $H_0$ is accepted or rejected on one diagonal, and the number of cases where one group passed the test while the other one failed on the other diagonal. The data arrangement in this case is shown in \cref{Table:Contingency Table - paired data} and we compare $\pi_{ij}$ with the \emph{most efficient} physics configuration $\hat{\pi}_i=\pi_{ik}$ as determined in the previous stage (see \cref{sec:statistic:gof}).
The data arrangement in $2\times 2$ contingency table for unpaired data is illustrated in \cref{Table:Contingency Table - unpaired data}. Here the number of rejected and accepted cases is marginalised over both $\pi_{ik}$ and the production cut value $c_j$, and is identical to $n_{i,\mathrm{rej}}$, $n_{i,\mathrm{acc}}$ from the previous section.

\begin{table}[ht]
\centering
\caption{Data arrangement in a contingency table for paired data as obtained from physics configurations $\pi_j=\pi_{ij}(P_i,c_j)$ and $\pi_k=\pi_{ik}=(P_i,c_k)$ with the same physics constructor $P_i$ but different production cut values $c_j$, $c_k$. The null hypothesis $H_0$ is \emph{rejected} (rej.) if $\pi$ is not compatible with the reference physics configuration $\pi_\mathrm{ref}$, otherwise \emph{accepted} (acc.). The cell values contain the correlation with the most efficient physics configuration $\pi_k=\hat{\pi}$. For details, see text.}
\begin{tabular*}{\linewidth}{@{\extracolsep{\fill}}ccc}
\toprule
 & {$\pi_j$; $H_0$ acc.} & {$\pi_j$; $H_0$ rej.} \\ \midrule
{$\pi_k$; $H_0$ acc.}                        & $n_{j:\mathrm{acc}, k:\mathrm{acc}}$        & $n_{j:\mathrm{acc}, k:\mathrm{rej}}$        \\
{$\pi_k$; $H_0$ rej.}                        & $n_{j:\mathrm{rej}, k:\mathrm{acc}}$        & $n_{j:\mathrm{rej}, k:\mathrm{rej}}$        \\
\bottomrule
\end{tabular*}%
\label{Table:Contingency Table - paired data}
\end{table}

\begin{table}[ht]
\centering
\caption{Data arrangement in a contingency table for unpaired data as obtained from physics configurations $\pi_i=\sum_n (P_i,c_n)$ and $\pi_j=\sum_n (P_j,c_n)$ with different physics constructors $P_i$, $P_j$ and marginalised production cut values $c_n$. The null hypothesis $H_0$ is \emph{rejected} (rej.) if $\pi$ is not compatible with the reference physics configuration $\pi_\mathrm{ref}$, otherwise \emph{accepted} (acc.). For details, see text.}
\begin{tabular*}{\linewidth}{@{\extracolsep{\fill}}ccc}
\toprule
                  & $\pi_i$ & $\pi_j$ \\ \midrule
$H_0$ acc.    & $n_{i,\mathrm{acc}}$   & $n_{j,\mathrm{acc}}$        \\
$H_0$ rej.    & $n_{i,\mathrm{rej}}$ & $n_{j,\mathrm{rej}}$        \\
\bottomrule
\end{tabular*}%
\label{Table:Contingency Table - unpaired data}
\end{table}

Test statistics are used to estimate whether there is a significant difference between physics configurations with respect to the obtained frequencies: 
Pearson's $\chi^2$ test of independence and Fisher’s exact test \cite{fisher} for unpaired data (i.e.\ physics configurations with different physics constructors), and McNemar’s test \cite{McNemar} for paired data (i.e.\ physics configurations with the same physics constructor but different values for the production cut). Results are also given for Yates's correction \cite{Yates1934}, which is adjusted for the Pearson's $\chi^2$ test, and prevents overestimation of statistical significance for contingency tables involving small numbers, but tends to yield a more conservative result.

The null hypothesis $H_0$ for these tests is that there is no relationship between the physics configurations, and if the $p$-value is lower than $\alpha$ we reject $H_0$ and conclude that there is a significant difference between the physics configurations. This means that the ratio $n_{i,\mathrm{acc}}/n_{i,\mathrm{rej}}$ of test cases for which $H_0$ is accepted or rejected, respectively, for a given physics configuration $\pi_{ij}$ differs from other physics configurations.

\begin{figure*}[ht]
\centering
\includegraphics[width=0.9\textwidth]{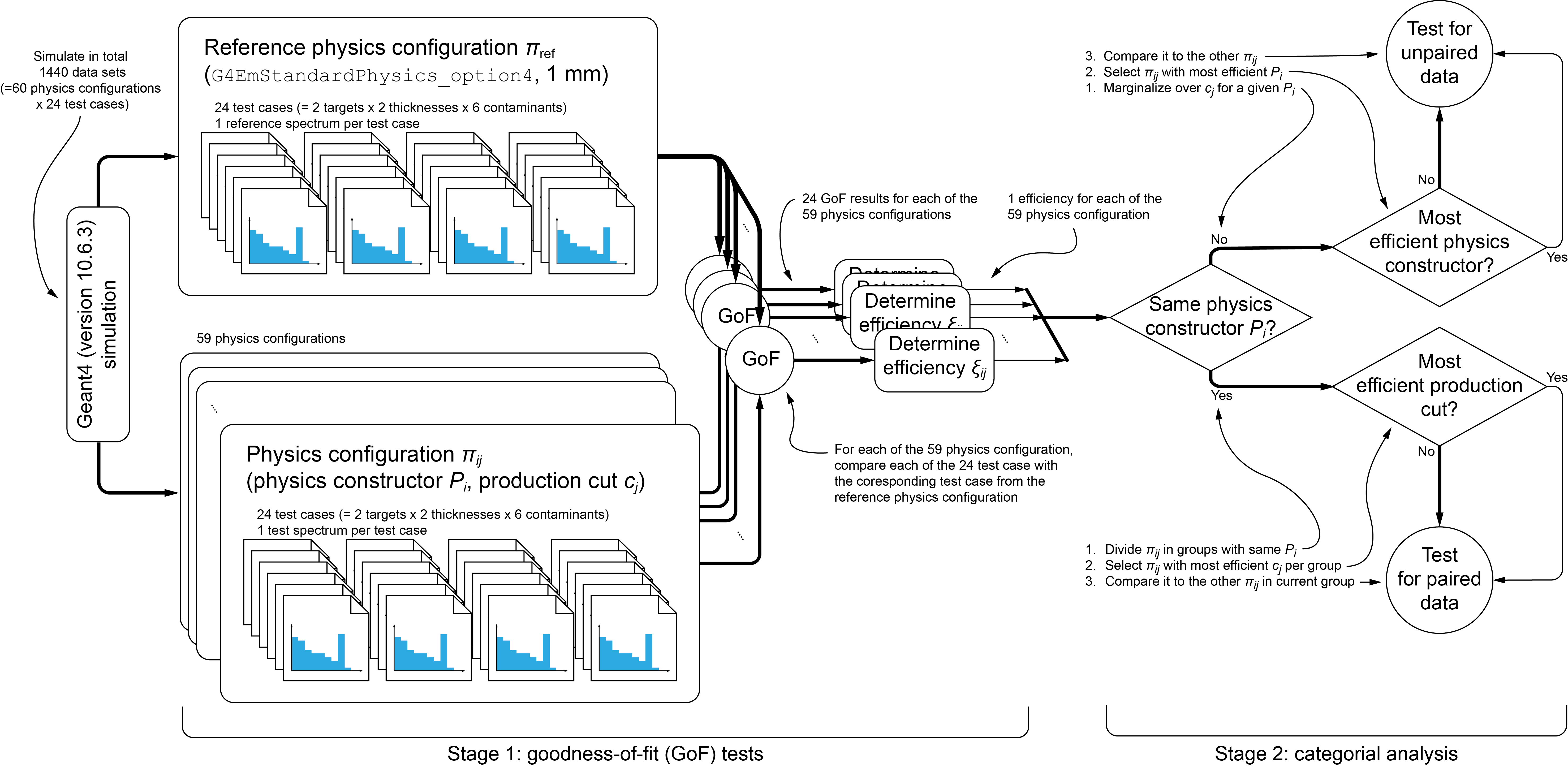}
\caption{Illustration of the applied statistical analysis. For paired data, only McNemar's test was suitable in stage 2; however, other parts of the workflow were repeated with several statistical tests to avoid systematic biases: For unpaired data both Pearson's $\chi^2$ test of independence and Fisher's exact test was used in stage 2; stage 1 of the workflow was always repeated for $\chi^2$ test, Kolmogorov-Smirnov test, and Anderson-Darling test as goodness-of-fit test. For details, see text.}
\label{fig:scheme}
\end{figure*}

The data analysis was conducted with code based on ROOT \cite{Brun1997} and R \cite{R}, and is schematically illustrated in Figure~\ref{fig:scheme}. 

\section{\label{sec:results}Obtained Compatibilities}

Applying the statistical methodology outlined in \cref{sec:statistic} to the 60 physics configurations (see \cref{sec:physicslists}) in 24 test cases (see \cref{sec:configurations}), we discuss the results under three aspects:
in \cref{sec:results:eDep} we show examples of agreement or discrepancy of the simulated spectra for selected test cases, in \cref{sec:results:gof} we give the efficiency of each physics configurations for all test cases, and in \cref{sec:results:cat} we determine if the physics configurations differ significantly from each other in terms of their efficiencies.

\subsection{Qualitative Compatibility with Reference Spectrum}\label{sec:results:eDep}
 
\begin{figure*}[ht]
\centering
\subfloat[Subfigure 1][]{
\includegraphics[width=0.45\textwidth]{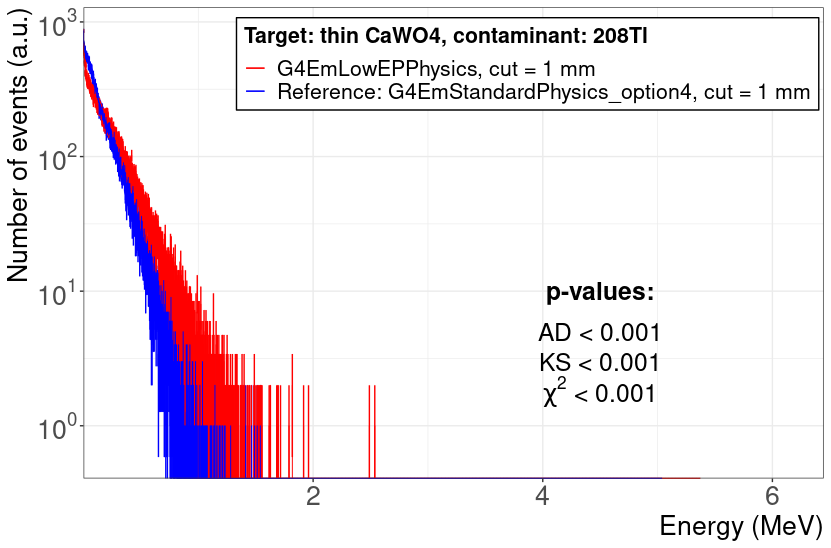}
\label{fig:208Tl_CaWO4_EP}}
\subfloat[Subfigure 2][]{
\includegraphics[width=0.45\textwidth]{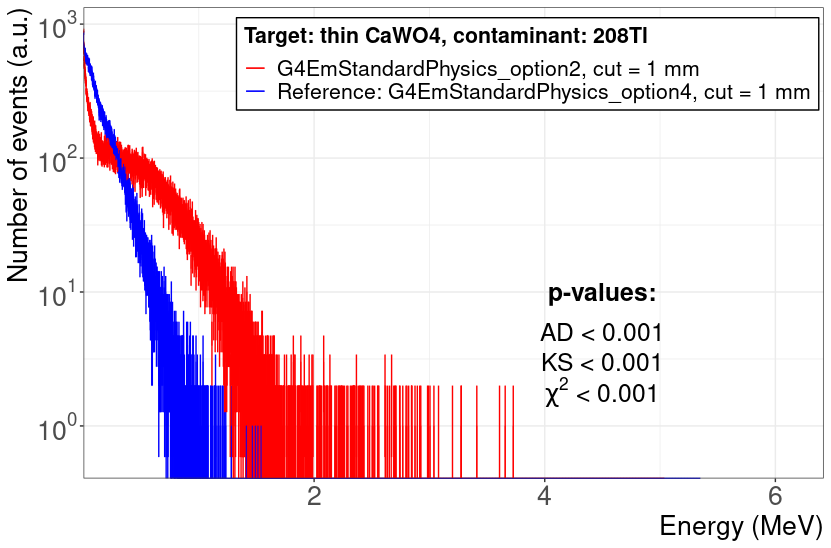}
\label{fig:208Tl_CaWO4_option2}}\\
\subfloat[Subfigure 4][]{
\includegraphics[width=0.45\textwidth]{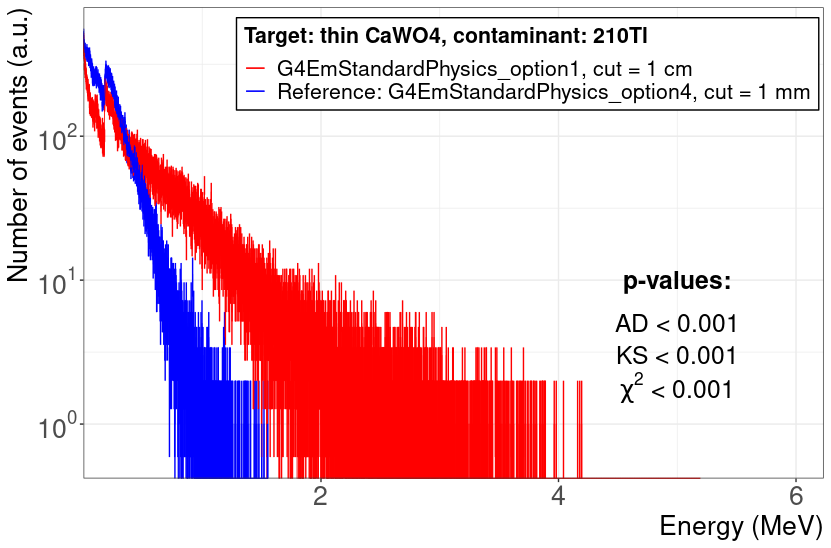}
\label{fig:210Tl_CaWO4_option1b}}
\subfloat[Subfigure 3][]{
\includegraphics[width=0.45\textwidth]{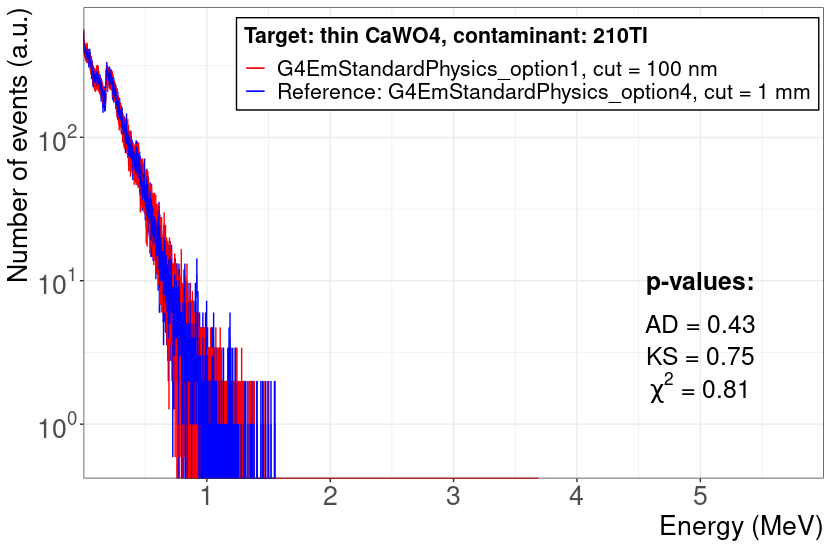}
\label{fig:210Tl_CaWO4_option1}}
\caption{Impact of physics constructor and production cut on the comparability between spectra of different physics configurations (\emph{red} histogram) and the reference spectrum (\emph{blue} histogram) for the example of total energy deposition by contaminants with large $Q$-values (\ce{^{208}Tl},\ce{^{210}Tl}) in a \qty{100}{\um}-thick \cawo{} target: (\subref{fig:208Tl_CaWO4_EP}, \subref{fig:208Tl_CaWO4_option2}) for the same production cut value but different physics constructors; (\subref{fig:210Tl_CaWO4_option1}, \subref{fig:210Tl_CaWO4_option1b}) for the same physics constructor but different production cut values. $p$-values are given for Anderson-Darling (AD), Kolmogorov-Smirnov (KS), and $\chi^2$ tests.}
\label{fig:spectra}
\end{figure*}

\begin{figure*}[ht]
\centering
\subfloat[Subfigure 5][]{
\includegraphics[width=0.45\textwidth]{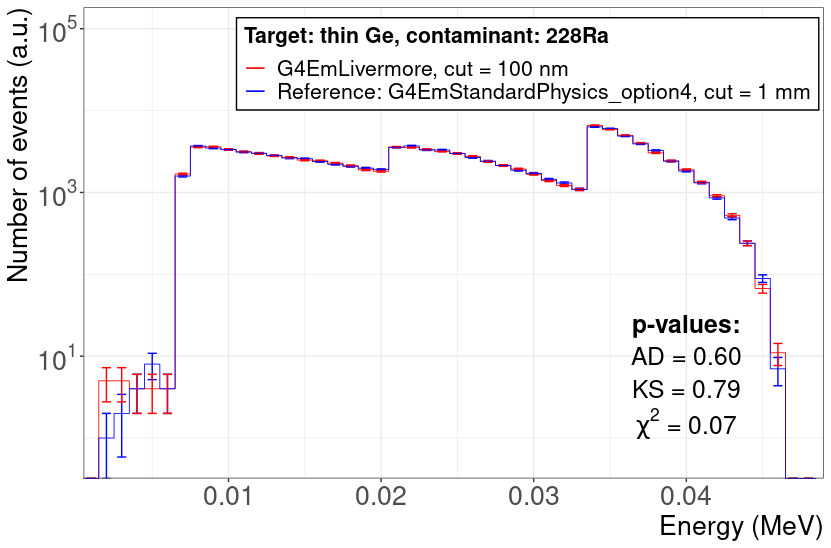}
\label{fig:228Ra_Ge_Livermore}}
\subfloat[Subfigure 6][]{
\includegraphics[width=0.45\textwidth]{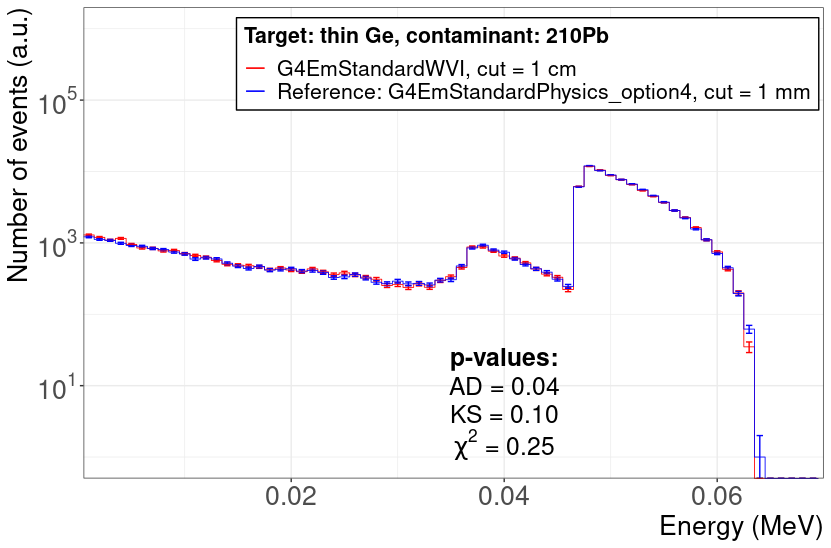}
\label{fig:210Pb_Ge_WVI}}\\
\subfloat[Subfigure 7][]{
\includegraphics[width=0.45\textwidth]{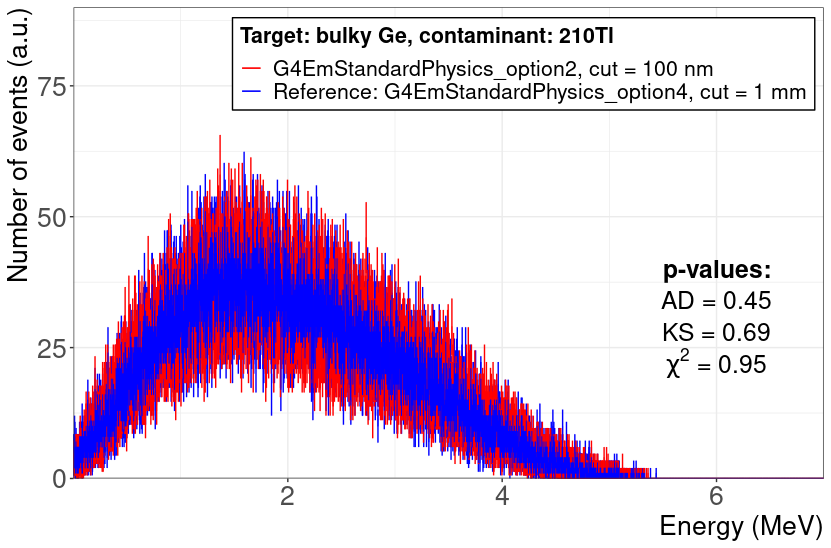}
\label{fig:210Tl_Ge_option2}}
\subfloat[Subfigure 8][]{
\includegraphics[width=0.45\textwidth]{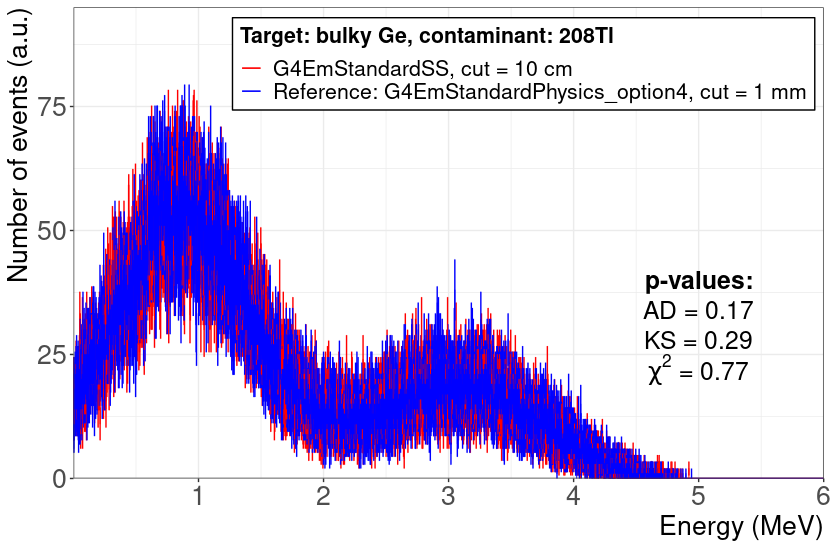}\label{fig:208Tl_Ge_SS}}
\caption{Examples of spectra for configurations (\emph{red} histogram), i.e.\ pairs of physics constructor and production cut value, that are compatible with the reference spectrum (\emph{blue} histogram) for various contaminants in \emph{thin} (\qty{100}{\um}-thick, \subref{fig:228Ra_Ge_Livermore}, \subref{fig:210Pb_Ge_WVI}) and \emph{bulky} (\qty{64}{\mm}-thick, \subref{fig:210Tl_Ge_option2}, \subref{fig:208Tl_Ge_SS}) \ce{Ge} targets. $p$-values are given for Anderson-Darling (AD), Kolmogorov-Smirnov (KS), and $\chi^2$ tests.}
\label{fig:spectra2}
\end{figure*}

To illustrate the compatibility and discrepancies between different physics configurations and the reference physics configuration, \cref{fig:spectra,fig:spectra2} give some examples for selected test cases. Each figure shows the spectra of total energy deposition per single event, i.e.\ the decay of a radioactive contaminant in the target material; each plot includes the $p$-values of GoF tests performed to study the statistical significance of the compatibility of  the test spectrum with the reference spectrum.

\Cref{fig:spectra} shows test cases with a high $Q$-value contaminant in a thin \cawo{} crystal. Due to a thickness of only \qty{100}{\um}, the deposited energy is significantly less than the full $Q$-value, cf.\ \cref{tab:contaminants}, as some fraction of the decay products leaks out of the target. The interplay between $Q$-value and target thickness is shown in \cref{fig:spectra2} for thin and bulky \ce{Ge} targets, and low and high $Q$-value contaminants. Also, thin targets can absorb the full $Q$-value if it is low enough (\cref{fig:228Ra_Ge_Livermore,fig:210Pb_Ge_WVI}). Similar, also high $Q$-values can be fully absorbed by the target if it is thick enough (\cref{fig:210Tl_Ge_option2,fig:208Tl_Ge_SS}).

\Cref{fig:208Tl_CaWO4_EP,fig:208Tl_CaWO4_option2} demonstrate the impact of different physics constructors, here we chose arbitrarily \texttt{G4EmLow\-EP\-Physics} and \texttt{G4EmStandardPhysics\_option2} as example, while keeping the reference value of \qty{1}{\mm} for the production cut: Both physics constructors cause a harder spectrum compared to our reference physics constructor, \texttt{G4EmStandardPhysics\_option4}, indicating a reduced leakage of decay products, which results in a statistical significant discrepancy, see the $p$-values in the plots. However, different physics constructors can also be statistical compatible, as shown in \cref{fig:228Ra_Ge_Livermore,fig:210Pb_Ge_WVI}.

\Cref{fig:210Tl_CaWO4_option1,fig:210Tl_CaWO4_option1b} demonstrate the impact of different production cut values while keeping the same physics constructor, here \texttt{G4EmStandardPhysics\_option1} as example: a production cut value of \qty{1}{cm}, i.e.\ \num{e4} times larger than the target thickness of \qty{100}{\um}, results in a strong discrepancy with respect to the reference spectrum, see \cref{fig:210Tl_CaWO4_option1b}. This can be remedied with a production cut value smaller than the target thickness: For \qty{100}{\nm}, the spectrum obtained with \texttt{G4EmStandardPhysics\_option1} is statistical compatible with the reference spectrum, see also the $p$-values in \cref{fig:210Tl_CaWO4_option1}. However, albeit it is prudent to set a production cut value smaller than the target dimensions, there is no general rule that the production cut \emph{has to be} smaller than the target dimensions as demonstrated in \cref{fig:210Pb_Ge_WVI,fig:208Tl_Ge_SS}: Here, the production cut is between a factor \num{\approx1.5} (\cref{fig:208Tl_Ge_SS}) and \num{e4} (\cref{fig:210Pb_Ge_WVI}) larger than the target thickness, but the test spectrum is still compatible with the reference spectrum based on the obtained $p$-values.

\Cref{fig:228Ra_Ge_Livermore,fig:210Pb_Ge_WVI} also show the occurrence of inconsistent results between different GoF tests. For \cref{fig:228Ra_Ge_Livermore}, the $\chi^2$ test yields a $p$-value considerably lower than the ones from KS and AD tests. For figure \cref{fig:210Pb_Ge_WVI}, the AD test even rejects a compatibility at \qty{5}{\percent} significance level, whereas KS and $\chi^2$ tests accept compatibility. This example affirm our decision to not rely on only one GoF test, but consider several.

\subsection{Efficiencies of Physics Configuration}\label{sec:results:gof}

Based on the GoF tests that were performed for each test case (see previous \cref{sec:results:eDep}), the efficiency of a given physics configuration was determined. In test cases where the resulting spectra are not continuous, only the $\chi^2$ test was performed because the AD and KS tests are only exact for continuous variables. This is e.g.\ the case for the \textalpha{} decaying contaminants \ce{^{211}Bi} and \ce{^{234}U} where the simulated spectrum consist of a very sharp peak near the $Q$-value with empty bins otherwise. \Cref{sec:results:gof:total} reports the total efficiencies, \cref{sec:results:gof:thickness} per target thickness, and \cref{sec:results:gof:material} per target material.

\subsubsection{Total Efficiencies}\label{sec:results:gof:total}

\begin{figure*}[ht]
\centering
\subfloat[Subfigure 1][]{
\includegraphics[width=0.5\textwidth]{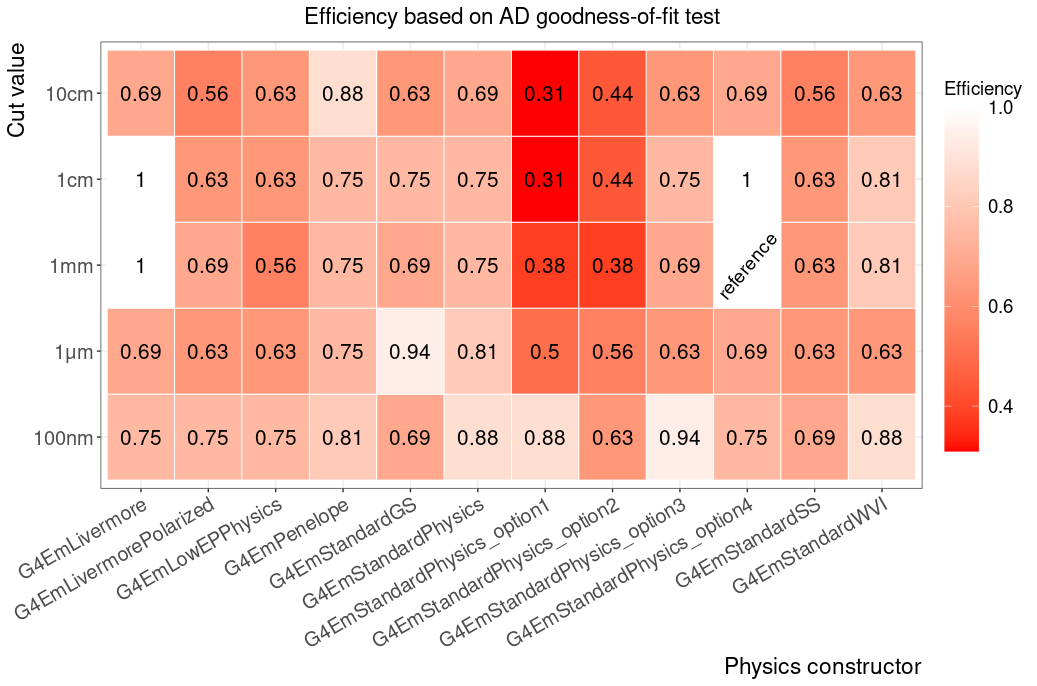}
\label{fig:eff_AD}}
\subfloat[Subfigure 2][]{
\includegraphics[width=0.5\textwidth]{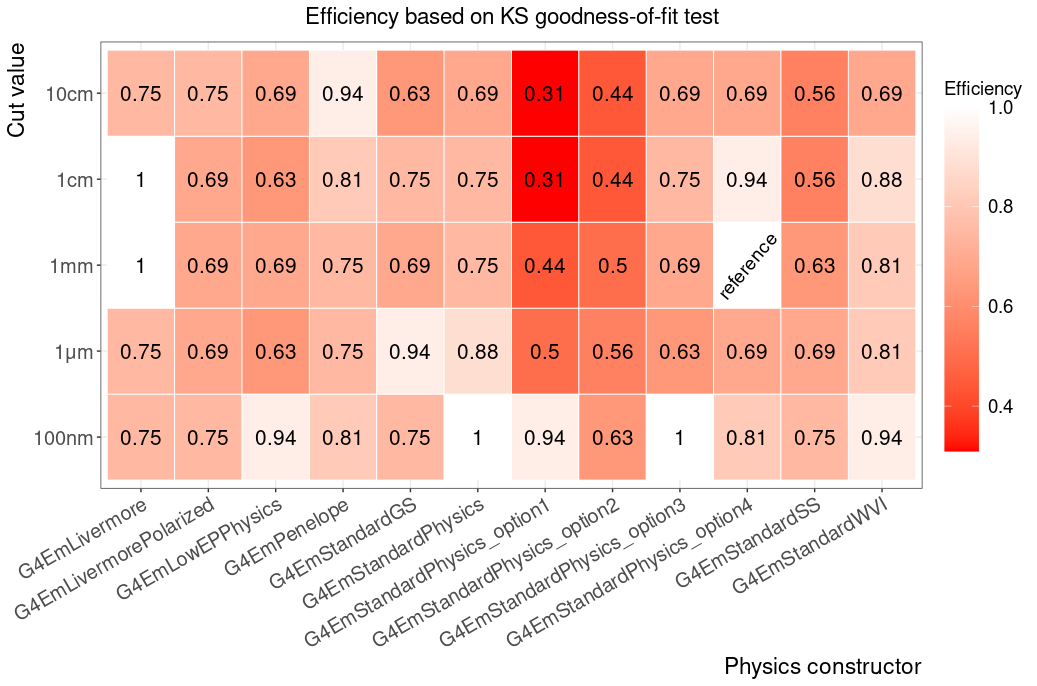}
\label{fig:eff_KS}}\\
\subfloat[Subfigure 3][]{
\includegraphics[width=0.5\textwidth]{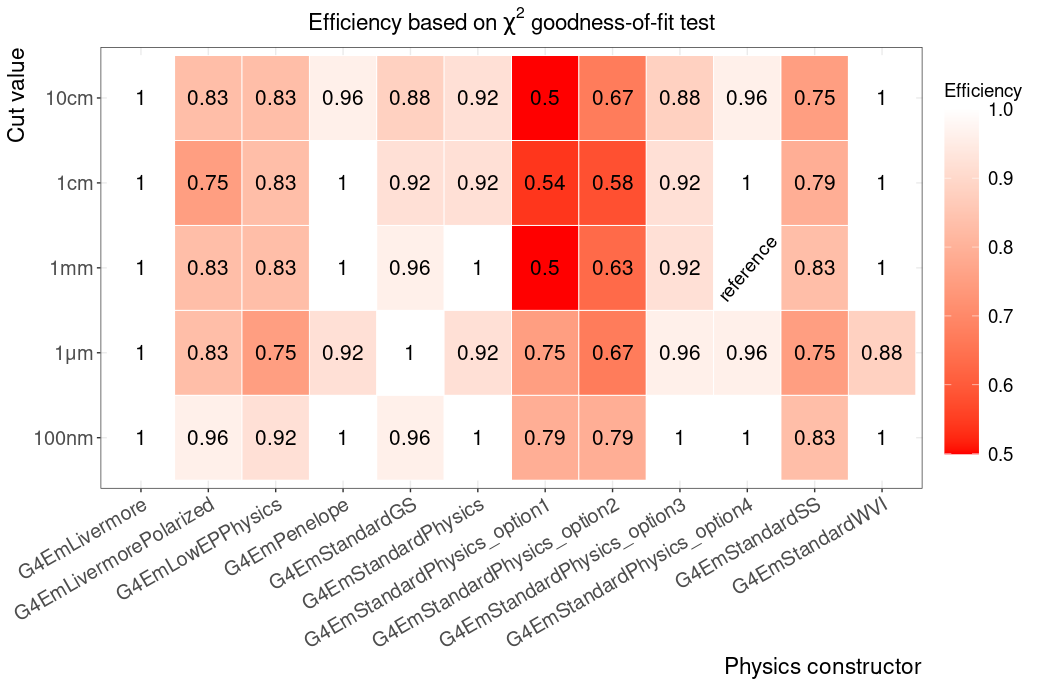}
\label{fig:eff_Chi2}}
\subfloat[Subfigure 4][]{
\includegraphics[width=0.5\textwidth]{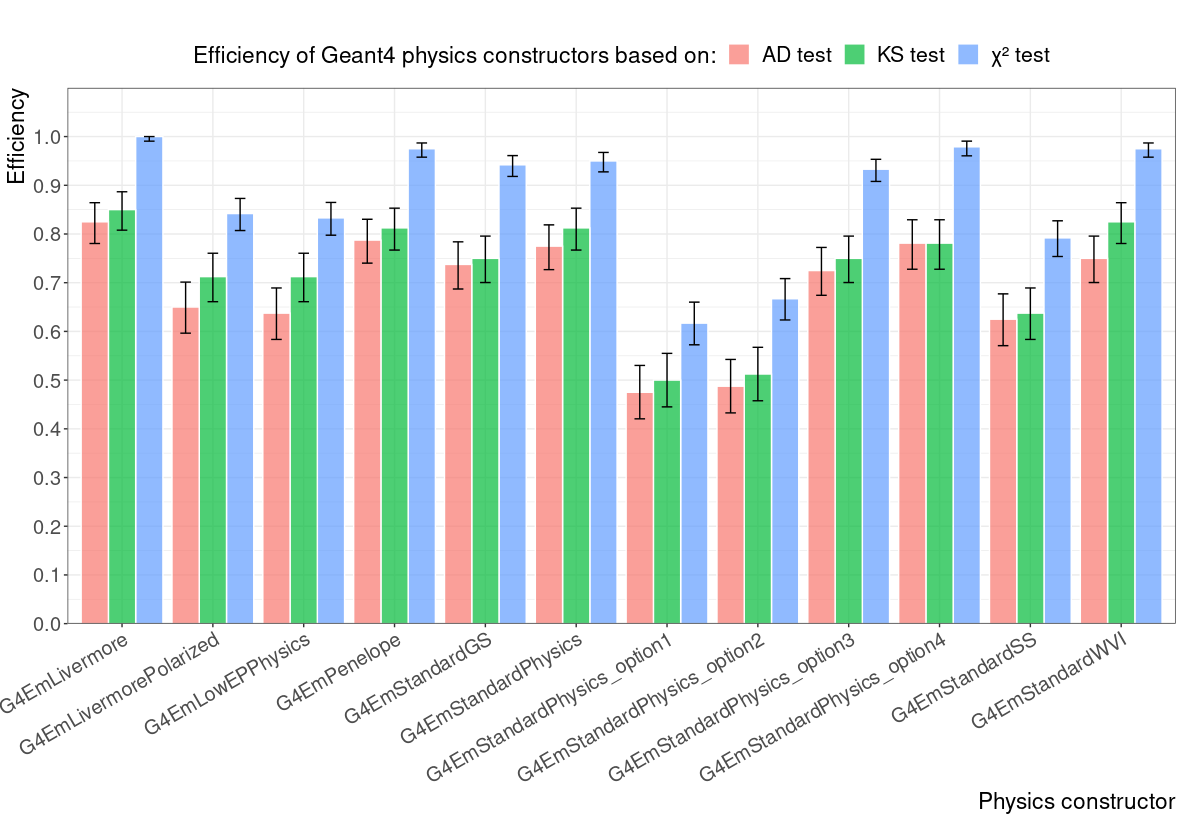}
\label{fig:total_eff}}
\caption{Efficiencies of \geant{} physics configurations, i.e.\ pairs of physics constructor and production cut value, for different goodness-of-fit tests: (\subref{fig:eff_AD}) Anderson-Darling (AD), (\subref{fig:eff_KS}) Kolmogorov-Smirnov (KS), and (\subref{fig:eff_Chi2}) $\chi^2$. The total efficiencies of physics constructors marginalised over the production cut value is shown in (\subref{fig:total_eff}).}
\label{fig:Efficiencies}
\end{figure*}

Marginalising over target material and target thickness, we obtain the efficiencies shown in \cref{fig:Efficiencies}. 

\Cref{fig:eff_AD,fig:eff_KS,fig:eff_Chi2} show the dependency of the efficiencies on the value for the production cut but different GoF tests;  uncertainties of these efficiencies are shown in \cref{fig:uncertainties}. Simulations using lower production cut value appear systematically more compatible in comparison with those using higher cuts. Sensitivity to the value of the production cut is observed particularly in case of \texttt{G4EmStandardPhysics\_option1} where all physics configurations perform noticeably less accurately than the one using the lowest cut value of \qty{100}{\nm}.  The special sensitivity of \texttt{G4EmStandardPhysics\_option1} is explainable, as it is the only one of the investigated EM constructors that applies the production cut to all EM interactions. These qualitative observations are quantified through categorical analysis in \cref{sec:results:cat:cut}.

The bar plot in \cref{fig:total_eff} shows the total efficiencies of \geant{} physics constructors marginalised over the value of the production cut. It demonstrates an apparent difference in compatibility observed across the \geant{} constructors: The highest efficiency of \qty{100}{\%} was achieved by \texttt{G4EmLivermore} for which the $\chi^2$ test fails to reject the hypothesis of compatibility in all cases. The worst compatibility with the reference \geant{} physics constructor is obtained by the physics constructors \texttt{G4EmStandardPhysics\_option1} and \texttt{G4EmStandardPhysics\_option2}, both resulting in efficiencies equal or lower than \qty{67}{\%} for the $\chi^2$ test. The poorer compatibility for our test cases, relatively small detector volumes, is expected as both constructors are optimised for high performance simulations of the relatively large volumes of LHC detectors (see \cref{table:constructor_comparison}). These results are qualitatively confirmed by the AD and KS tests.

The most efficient physics constructor $\hat{P}$ is \texttt{G4EmLivermore} which has an efficiency even higher than the reference physics constructor \texttt{G4EmStandardPhysics\_option4}. As shown in \Cref{fig:eff_AD,fig:eff_KS,fig:eff_Chi2}, the efficiency of \texttt{G4EmStandard\linebreak{}Physics\_option4} depends stronger on the cut value than it is the case for \texttt{G4EmLivermore}. This means that for certain test cases, the spectrum obtained with \texttt{G4Em\-Standard\-Physics\_option4} and the reference cut value of \qty{1}{\mm} differs statistical significantly from the spectrum obtained with the same physics constructor but a different cut value. Whereas for \texttt{G4EmLivermore} spectra obtained with different cut values differs less. The exact efficiency depends on the used GoF test, i.e.\ the $\chi^2$ test assigns a \qty{100}{\percent}-efficiency independent of the cut value, but the qualitative behaviour is supported also by AD\footnote{We note that the efficiency determined by the AD test shown in \cref{fig:total_eff} is less for \texttt{G4EmStandardPhysics\_option4} then for \texttt{G4EmLivermore} albeit \cref{fig:eff_AD} shows nearly the same efficiencies per cut value. This is because the test case with the reference cut value of \qty{1}{\mm} is counted for \texttt{G4EmLivermore} but not for the reference constructor \texttt{G4EmStandardPhysics\_option4} - in the latter case the GoF would have to compare it to itself and hence it is excluded.} and KS test.

\subsubsection{Per Target Thickness}\label{sec:results:gof:thickness}

\begin{figure*}[ht]
\centering
\subfloat[Subfigure 1][]{
\includegraphics[width=0.5\textwidth]{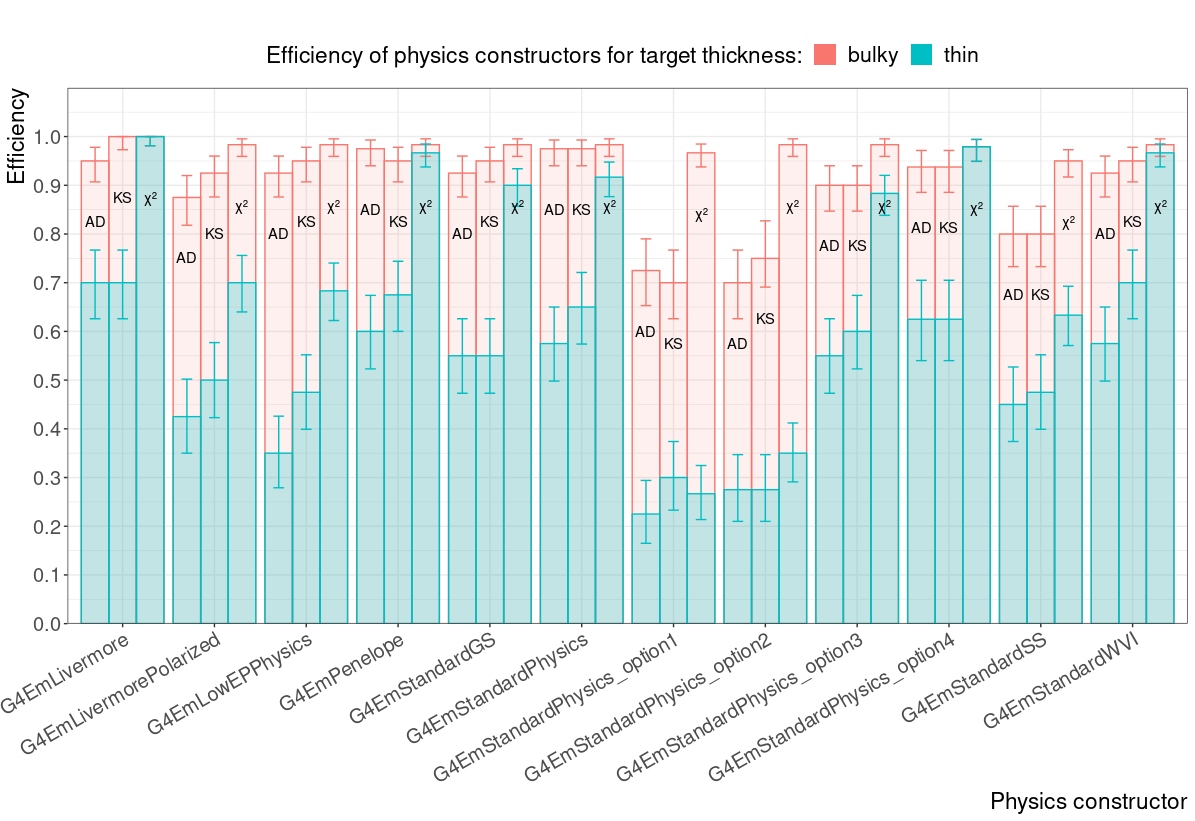}
\label{fig:partial_eff_thickness}}
\subfloat[Subfigure 2][]{
\includegraphics[width=0.5\textwidth]{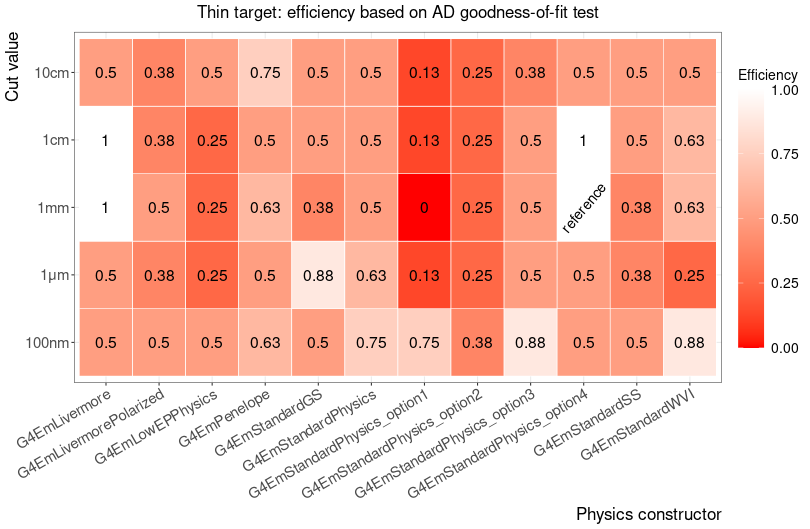}
\label{fig:thin_AD}}
\qquad
\subfloat[Subfigure 3][]{
\includegraphics[width=0.5\textwidth]{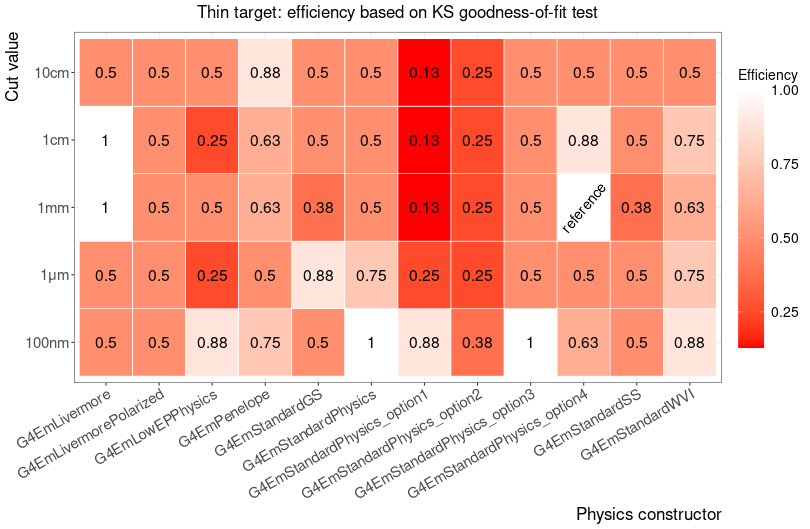}
\label{fig:thin_KS}}
\subfloat[Subfigure 4][]{
\includegraphics[width=0.5\textwidth]{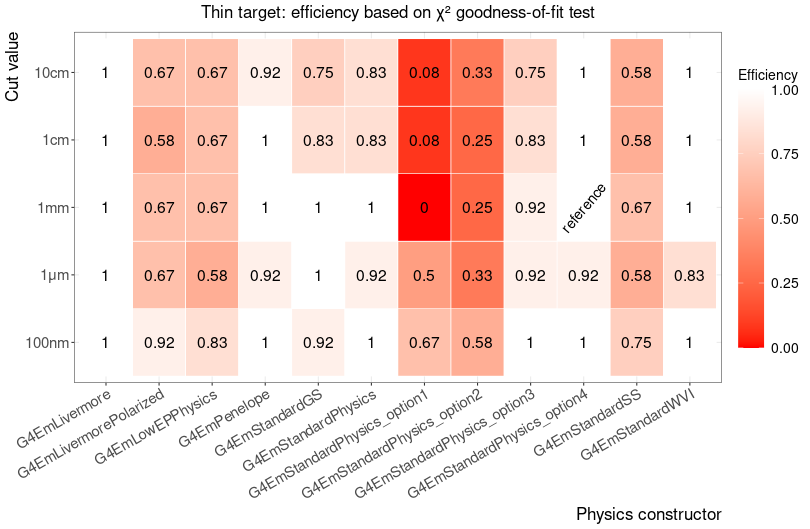}
\label{fig:thin_Chi2}}
\caption{The total efficiencies of \geant{} physics constructors marginalised over the production cut value and target material is shown in (\subref{fig:partial_eff_thickness}) for bulky (\qty{64}{\mm}-thickness, \emph{red}) and thin (\qty{100}{\um}-thickness, \emph{blue}) targets; bar plots are not stacked. The efficiencies for different physics configurations, i.e.\ pairs of physics constructor and production cut value, are shown for different goodness-of-fit tests: (\subref{fig:thin_AD}) Anderson-Darling (AD), (\subref{fig:thin_KS}) Kolmogorov-Smirnov (KS), and (\subref{fig:thin_Chi2}) $\chi^2$.}
\label{fig:EfficienciesThickness}
\end{figure*}

The efficiencies separated by target thickness, bulky and thin, but marginalised over production cut value and target material are shown in \cref{fig:EfficienciesThickness}. As indicated in \cref{fig:partial_eff_thickness}, the efficiency for \emph{thin} targets varied stronger than for \emph{bulky} targets. This seems plausible, as the simulation of energy deposition in thin targets is more sensitive to individual particle interactions and hence to the details of the physics models used in a given physics constructor and the applied production cut value. \Cref{fig:thin_AD,fig:thin_KS,fig:thin_Chi2} show that the efficiencies for most physics constructors increase with decreasing production cut, i.e.\ a more fine-grained consideration of secondary particle tracking. However, similar to our observation in \cref{sec:results:eDep}, a smaller production cut value does not always improve the efficiency. 

\subsubsection{Per Target Material}\label{sec:results:gof:material}

\begin{figure}[ht]
    \centering
    \includegraphics[width=0.95\linewidth]{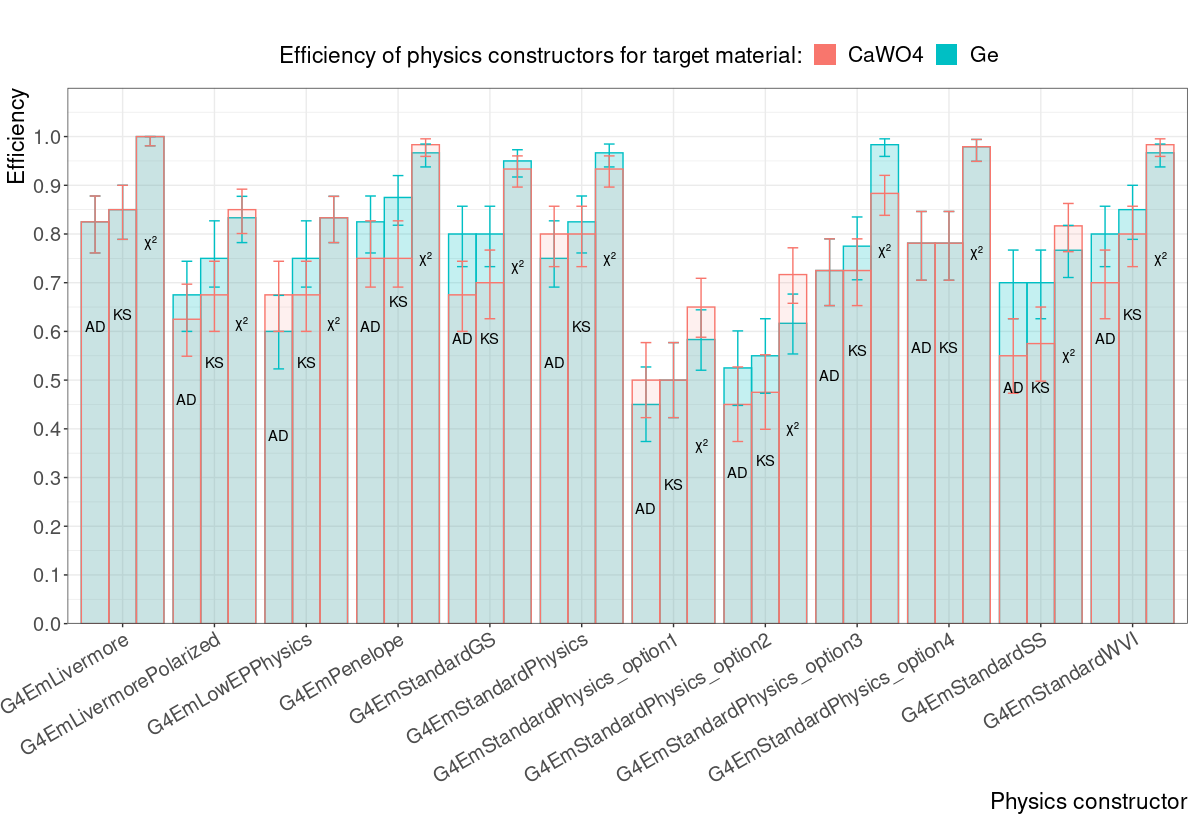}
    \caption{The total efficiencies of \geant{} physics constructors marginalised over the production cut value and target thickness is shown for \cawo{} (\emph{red}) and \ce{Ge} (\emph{blue}) targets; bar plots are not stacked. The efficiencies are shown for different goodness-of-fit tests: Anderson-Darling (AD), Kolmogorov-Smirnov (KS), and $\chi^2$.}
    \label{fig:partial_eff_material}
\end{figure}

The total efficiencies separated by target material but marginalised over production cut value and target thickness are shown in \cref{fig:partial_eff_material}. Within the uncertainties, all physics constructors show the same efficiency with respect to the target materials \cawo{} and \ce{Ge}.

\subsection{Compatibility of Different Physics Configurations}\label{sec:results:cat}
Following \cref{sec:statistic:cat}, we analyse first the impact of different production cuts for a given physics constructor on the compatibility in \cref{sec:results:cat:cut} and then marginalise over the production cut to compare the physics constructors to each other in \cref{sec:results:cat:const}. Because the GoF tests do not produce consistent outcomes regarding the rejection of the null hypothesis of compatibility, we report also the results of the categorical analyses for all three GoF tests. 

\subsubsection{Per Production Cut}\label{sec:results:cat:cut}
For each physics constructor $P_i$, the production cut value $\hat{c_i}$ that constitutes the configuration $\hat{\pi}_i=(P_i,\hat{c_i})$ with the highest efficiency was determined according to the GoF results shown in \cref{fig:eff_AD,fig:eff_KS,fig:eff_Chi2}. \Cref{table:RelatedDataResults} summarises the $p$-values of McNemar’s test, comparing the compatibility of physics configurations $\pi_{ij}$ with different production cuts $c_j$ with $\hat{\pi}_i$ that produce the highest efficiency for a given physics constructor $P_i$. Configurations with equivalent efficiencies are marked with a line \enquote{—–} \footnote{In this case, there is no difference in proportions of counts across the categories and therefore the test for a significant difference between the data was not performed.}.

\begin{table*}[ht]
\caption{McNemar's $p$-values for a given physics configuration $\pi_{ij}=(P_i,c_j)$, i.e. pairs of physics constructor $P_i$ and production cut value $c_j$, compared to the most efficient configuration $\hat{\pi}_i$ for those physic constructors. Results are given for all three goodness-of-fit tests that were used to determine the efficiency with respect to the reference configuration $\pi_\mathrm{ref}$: Anderson-Darling (AD), Kolmogorov-Smirnov (KS), and $\chi^2$. A $p$-value less than the significance level $\alpha=\qty{5}{\%}$ indicates a difference between $\pi_{ij}$ and $\hat{\pi}_i$ that is significant at the \qty{95}{\%} confidence level; a $p$-value greater than $\alpha$ means there is no significant difference. Configurations with equivalent efficiencies are marked with a line \enquote{—–}. For details, see text.}
\resizebox{\textwidth}{!}{%
\begin{tabular}{lccccccccccccccc}
\toprule
 & \multicolumn{15}{c}{Production cut value $c_j$}\\\cmidrule{2-16}
              & \multicolumn{3}{c}{\qty{100}{\nm}}                                                  & \multicolumn{3}{c}{\qty{1}{\um}}                               & \multicolumn{3}{c}{\qty{1}{\mm}}                                              & \multicolumn{3}{c}{\qty{1}{\cm}}                                  & \multicolumn{3}{c}{\qty{10}{\cm}}                                             \\\cmidrule(lr){2-4} \cmidrule(lr){5-7} \cmidrule(lr){8-10} \cmidrule(lr){11-13} \cmidrule(lr){14-16}
Physics constructor $P_i$                 & $\chi^2$                   & KS                        & AD                        & $\chi^2$ & KS                        & AD                        & $\chi^2$             & KS                        & AD                        & $\chi^2$ & KS                        & AD                        & $\chi^2$             & KS                        & AD                        \\ \midrule
\texttt{G4EmLivermore}                & -----                      & 0.134                      & 0.134                      & -----    & 0.134                      & 0.074                      & -----                & $\hat{\pi}_i$ & $\hat{\pi}_i$ & -----    & -----                      & -----                      & -----                & 0.134                      & 0.074                      \\
\texttt{G4EmLivermorePolarized}       & $\hat{\pi}_i$ & $\hat{\pi}_i$ & $\hat{\pi}_i$ & 0.371    & 1.000                      & 0.480                      & 0.371                & 1.000                      & 1.000                      & 0.131    & 1.000                      & 0.480                      & 0.371                & -----                      & 0.24821                    \\
\texttt{G4EmLowEPPhysics}             & $\hat{\pi}_i$ & $\hat{\pi}_i$ & $\hat{\pi}_i$ & 0.289    & 0.074                      & 0.480                      & 0.683                & 0.134                      & 0.248                      & 0.683    & 0.074                      & 0.480                      & 0.683                & 0.134                      & 0.617                      \\
\texttt{G4EmPenelope}                 & $\hat{\pi}_i$ & 0.617                      & 1.000                      & 0.480    & 0.248                      & 0.480                      & -----                & 0.248                      & 0.480                      & -----    & 0.480                      & 0.480                      & 1.000                & $\hat{\pi}_i$ & $\hat{\pi}_i$ \\
\texttt{G4EmStandardGS}               & 1.000                      & 0.248                      & 0.134                      & $\hat{\pi}_i$      & $\hat{\pi}_i$ & $\hat{\pi}_i$ & 1.000                & 0.134                      & 0.134                      & 0.480    & 0.248                      & 0.248                      & 0.248                & 0.074                      & 0.074                      \\
\texttt{G4EmStandardPhysics}          & $\hat{\pi}_i$ & $\hat{\pi}_i$ & $\hat{\pi}_i$ & 0.480    & 0.480                      & 1.000                      & -----                & 0.134                      & 0.617                      & 0.480    & 0.134                      & 0.617                      & 0.480                & 0.074                      & 0.371                      \\
\texttt{G4EmStandardPhysics\_option1} & $\hat{\pi}_i$ & $\hat{\pi}_i$ & $\hat{\pi}_i$ & 1.000    & $<\alpha$       & $<\alpha$       & $<\alpha$ & $<\alpha$       & $<\alpha$       & 0.077    & $<\alpha$       & $<\alpha$       & $<\alpha$ & $<\alpha$       & $<\alpha$       \\
\texttt{G4EmStandardPhysics\_option2} & $\hat{\pi}_i$ & $\hat{\pi}_i$ & $\hat{\pi}_i$ & 0.371    & 1.000                      & 1.000                      & 0.134                & 0.480                      & 0.134                      & 0.074    & 0.248                      & 0.248                      & 0.371                & 0.248                      & 0.248                      \\
\texttt{G4EmStandardPhysics\_option3} & $\hat{\pi}_i$ & $\hat{\pi}_i$ & $\hat{\pi}_i$ & 1.000    & $<\alpha$       & 0.131                      & 0.480                & 0.074                      & 0.221                      & 0.480    & 0.134                      & 0.371                      & 0.248                & 0.074                      & 0.074                      \\
\texttt{G4EmStandardPhysics\_option4} & $\hat{\pi}_i$ & 0.617                      & 0.134                      & 1.000    & 0.134                      & 0.074                      & $\pi_\mathrm{ref}$            & $\pi_\mathrm{ref}$                  & $\pi_\mathrm{ref}$                  & -----    & $\hat{\pi}_i$ & $\hat{\pi}_i$ & 1.000                & 0.134                      & 0.074                      \\
\texttt{G4EmStandardSS}               & $\hat{\pi}_i$ & $\hat{\pi}_i$ & $\hat{\pi}_i$ & 0.480    & 1.000                      & 1.000                      & 0.480                & 0.480                      & 1.000                      & 1.000    & 0.248                      & 1.000                      & 0.617                & 0.248                      & 0.480                      \\
\texttt{G4EmStandardWVI}              & $\hat{\pi}_i$ & $\hat{\pi}_i$ & $\hat{\pi}_i$ & 0.248    & 0.480                      & 0.221                      & -----                & 0.617                      & 1.000                      & -----    & 0.480                      & 1.000                      & -----                & 0.134                      & 0.134                      \\ \bottomrule
\end{tabular}%
}
\label{table:RelatedDataResults}
\end{table*}

The categorical analysis performed found an unambiguous significant difference at the \qty{95}{\%} confidence level for the physics constructor \texttt{G4EmStandardPhysics\_option1}. Only for this case, analyses based on all three GoF tests found a significant difference. The significant difference of the physics configuration (\texttt{G4EmStandardPhysics\_option3}, \qty{1}{\micro\meter}) is only supported by the analysis based on the KS test. For all other cases, the hypothesis of statistically insignificant difference is accepted. This means that all production cut values $c_j$ for a given physics constructor $P_i$ show equivalent compatibility with respect to the best performing physics configuration $\hat{\pi}_i$.

\subsubsection{Per Physics Constructor}\label{sec:results:cat:const}
Marginalising over the production cut value $c_j$, the physics configuration $\pi_{ij}$ is reduced to the physics constructor $P_i$, and we obtain \cref{table:UnrelatedDataResults}. It shows $p$-values resulting from the comparison between the most efficient physics constructor $\hat{P}$, \texttt{G4EmLivermore}, and any other physics constructor $P_i$ based on the results of AD, KS and $\chi^2$ GoF tests. 

\begin{table*}[ht]
\caption{Yates' $p$-value and those from $\chi^2$ and Fisher's tests for a given physics constructor $P_i$ compared to the most efficient constructor \texttt{G4EmLivermore}. Results are given for
all three goodness-of-fit tests that were used to determine the efficiency with respect to the reference configuration: Anderson-Darling (AD), Kolmogorov-Smirnov (KS), and $\chi^2$. A $p$-value less than the significance level $\alpha=\qty{5}{\%}$ indicates a difference between $\pi_{ij}$ and $\hat{\pi}_i$ that is significant at the \qty{95}{\%} confidence level; a $p$-value greater than $\alpha$ means there is no significant difference. For details, see text.}
\resizebox{\textwidth}{!}{%
\begin{tabular}{lccccccccc}
\toprule
 & \multicolumn{9}{c}{Efficiencies determined via \ldots}\\ \cmidrule{2-9}
                             & \multicolumn{3}{c}{AD}                                  & \multicolumn{3}{c}{KS}                                  & \multicolumn{3}{c}{$\chi^2$}                            \\\cmidrule(lr){2-4}\cmidrule(lr){5-7}\cmidrule(lr){8-10}
Physics constructor $P_i$        & $\chi^2$ & Yates' $p$-value & Fisher’s test & $\chi^2$ & Yates' $p$-value & Fisher’s test & $\chi^2$ & Yates' $p$-value & Fisher’s test \\ \midrule
\texttt{G4EmLivermorePolarized}       & $<\alpha$      & $<\alpha$    & $<\alpha$         & $<\alpha$      & 0.056                   & 0.055                       & $<\alpha$      & $<\alpha$    & $<\alpha$         \\
\texttt{G4EmLowEPPhysics}             & $<\alpha$      & $<\alpha$    & $<\alpha$         & $<\alpha$      & 0.056                   & 0.055                       & $<\alpha$      & $<\alpha$    & $<\alpha$         \\
\texttt{G4EmPenelope}                 & 0.549                     & 0.689                   & 0.690                        & 0.527                     & 0.673                   & 0.674                       & 0.081                     & 0.245                   & 0.247                       \\
\texttt{G4EmStandardGS}               & 0.181                     & 0.251                   & 0.251                        & 0.114                     & 0.167                   & 0.166                        & $<\alpha$      & $<\alpha$    & $<\alpha$         \\
\texttt{G4EmStandardPhysics}          & 0.429                     & 0.553                   & 0.554                        & 0.527                     & 0.673                   & 0.674                        & $<\alpha$      & $<\alpha$    & $<\alpha$         \\
\texttt{G4EmStandardPhysics\_option1} & $<\alpha$      & $<\alpha$    & $<\alpha$         & $<\alpha$      & $<\alpha$    & $<\alpha$         & $<\alpha$      & $<\alpha$    & $<\alpha$         \\
\texttt{G4EmStandardPhysics\_option2} & $<\alpha$      & $<\alpha$    & $<\alpha$         & $<\alpha$      & $<\alpha$    & $<\alpha$         & $<\alpha$      & $<\alpha$    & $<\alpha$         \\
\texttt{G4EmStandardPhysics\_option3} & 0.130                     & 0.185                   & 0.185                        & 0.114                     & 0.167                   & 0.166                        & $<\alpha$      & $<\alpha$    & $<\alpha$         \\
\texttt{G4EmStandardPhysics\_option4} & 0.510                     & 0.655                   & 0.532                        & 0.286                     & 0.396                   & 0.384                        & 0.112                     & 0.382                   & 0.196                       \\
\texttt{G4EmStandardSS}               & $<\alpha$      & $<\alpha$    & $<\alpha$         & $<\alpha$      & $<\alpha$    & $<\alpha$         & $<\alpha$      & $<\alpha$    & $<\alpha$         \\
\texttt{G4EmStandardWVI}              & 0.246                     & 0.334                   & 0.334                        & 0.668                     & 0.830                   & 0.831                       & 0.081                     & 0.245                   & 0.247                       \\ \bottomrule
\end{tabular}%
}
\label{table:UnrelatedDataResults}
\end{table*}

For \texttt{G4EmStandardPhysics\_option1}, \texttt{G4EmStandardPhysics\_option2} and \texttt{G4EmStandardSS}, all analyses found significant differences at \qty{95}{\%} confidence level with respect to \texttt{G4EmLivermore}. Analyses based on $\chi^2$ test, either for determining the compatibility with the reference physics configuration or for conducting the unpaired categorical analysis, found also significant differences for \texttt{G4EmLivermorePolarized}, \texttt{G4EmLowEPPhysics}, \texttt{G4EmStandardPhysics\_option3}, and \texttt{G4EmStandardGS}, cf.\ \cref{table:UnrelatedDataResults}. 
No statistically significant difference in compatibility is observed between the remaining physics constructors: \texttt{G4EmWenzelWVI}, \texttt{G4EmPenelope}, \texttt{G4EmStandardPhysics\_option4}, and \texttt{G4EmLivermore}. We note that the latter three physics constructors share the same multiple-scattering model and the same model parameters, hence their compatibility may be no surprise.

Overall, the categorical analysis of contingency tables confirms the qualitatively visible difference in compatibility between most unpaired \geant{} physics constructors. Paired \geant{} physics configurations achieve statistically similar efficiencies, while some significant differences are visible only for two physics constructors: \texttt{G4EmStandardPhysics\_option1} and \texttt{G4EmStandardPhysics\_option3}.

\section{\label{sec:performance}Computing Performance of Physics Constructors}

Besides maximising the accuracy of simulations, one also wants to optimise the computational performance in terms of CPU time. Similar studies exist for high energy physics applications (e.g.\ \cite{Muskinja2020,Dotti2015}), but to our knowledge not for the energy range of natural radioactivity.

We run our simulations on two different computing clusters: \emph{The Max Planck Computing and Data Facility} (MPCDF, Munich)  \cite{MPCDF} and \emph{The Cloud Infrastructure Platform} (CLIP, Vienna) \cite{CLIP}. Part of our simulations were run on the Raven supercomputer of MPCDF, which consists of \num{1592} CPU compute nodes, \num{114624} CPU cores, and \qty{421}{\tera\byte} RAM memory. The remaining part of our simulation were run on the CLIP system which consist of \num{200} CPU compute nodes, \num{8000} CPU cores, and \qty{32}{\tera\byte} RAM memory. In both cases, we made no use of the GPU accelerators which are available on both clusters.

To estimate the impact of the different computing hardware on the observed simulation performance, we run a selected subset of them identically on both clusters: as an example of a physics constructor that requires high performances we choose \texttt{G4StandardSS}, and as an example for low requirements we choose the \texttt{G4Standard\_option1}, which is optimised for speed (see \cref{table:constructor_comparison}). As test cases we choose the contaminants \ce{^{208}Tl} and \ce{^{228}Ra}: the high $Q$-value of the former cause many secondary tracks, and hence requires relatively large computing power, contrary, the low $Q$-value of the latter cause only a few secondary tracks and hence require relatively lower computing power. These simulations were run several times on both clusters. The results indicate different computation resources between both clusters: in average, the simulation running on CLIP cluster consumes \qty{20}{\percent} less CPU time for same calculation as Raven. However, we observed also heterogeneous performances within a given cluster: at the Raven cluster, \qty{20}{\percent} of the jobs finish earlier than the average. We summarise these effects by assigning an uncertainty of \qty{30}{\percent} to the observed performances of the simulation conducted for this work.

Considering this uncertainty, we identified only two factors that significantly impacts our simulation performance: if the physics constructor uses a \emph{single-scattering} or \emph{hybrid} approach, or a \emph{multiple-scattering} approach to implement electron Coulomb scattering (see \cref{sec:physicslists}), and if the value for the production value drops below \qty{1}{\um}.

\begin{figure}[h]
    \centering
    \includegraphics[width=0.90\linewidth]{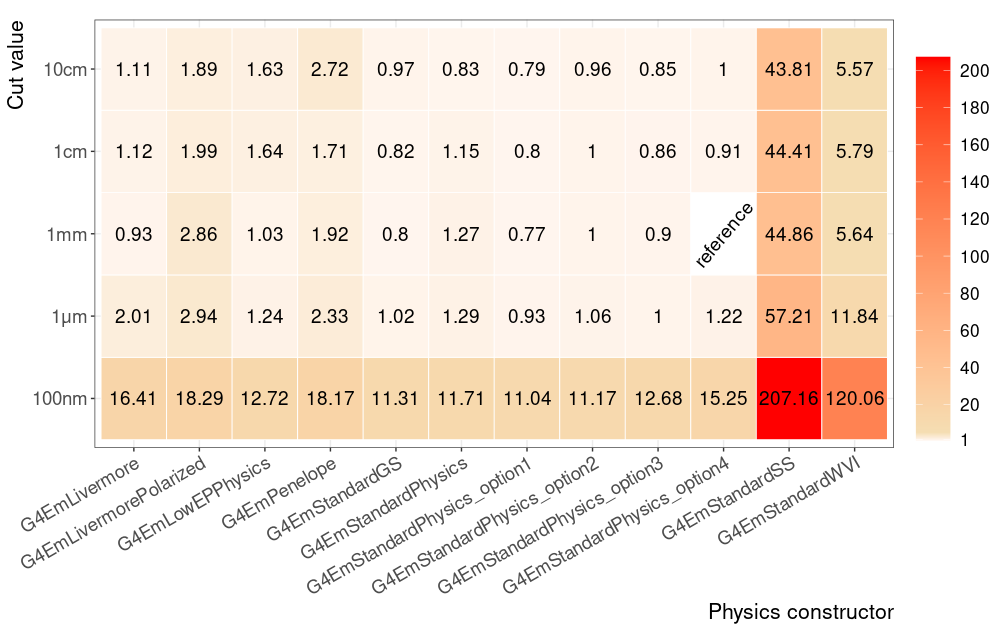}
    \caption{Average run time of \geant{} physics configurations relative to the reference configuration. Run time is plotted regardless of the target material and thickness.}
    \label{fig:run_time}
\end{figure}

\Cref{fig:run_time} shows the run time per physics configuration relative to our reference physics configuration and averaged over the test cases. Without surprise, physics configurations that are using physics constructors with a \emph{single-scattering} or \emph{hybrid} approach to implement electron Coulomb scattering, like \texttt{G4EmStandardSS} and \texttt{G4EmStandardWVI} (see \cref{sec:physicslists}), take up to $\mathcal{O}(\num{100})$ times longer. Physics configurations that rely on a \emph{multiple-scattering} approach, shows variations comparable with the uncertainty. With regard to the production cut value, the performance stays stable with regard to the uncertainty for all values \qty{>= 1}{\um}, but decreases by a factor of $\mathcal{O}(10)$ once it drops below.

\section{\label{sec:conclusion}Conclusion}

For \geant{} version 10.6.3, we investigated the impact of the used electromagnetic physics configurations on the simulated background spectrum in various test cases relevant for experiments searching for rare events. The test cases consist of six common radioactive contaminants (\ce{^{208}Tl}, \ce{^{210}Tl}, \ce{^{210}Pb}, \ce{^{211}Bi}, \ce{^{228}Ra}, \ce{^{234}U}), which are spanning an energy range from $\mathcal{O}(\qty{10}{\keV})$ to $\mathcal{O}(\qty{1}{\MeV})$ and covering \textalpha{}, \textbeta{} and \textgamma{} decays, in two common target materials (\ce{Ge}, \cawo{}) for bulky and thin geometries (target thicknesses of \qty{64}{\mm} and \qty{100}{\micro\meter}, respectively). The physics configurations consist of twelve \geant{} electromagnetic physics constructors, each configured with five values for the secondary particle production cut. In total, we simulated \num{1440} data sets.

We did not study the accuracy of the simulation with respect to specific experimental data, but the general systematic uncertainty. 
The impact of the electromagnetic physics configuration, i.e.\ the pair of physics constructor and production cut, on the total deposited energy inside the target is quantified as the statistical comparability with respect to a reference physics configuration. For this, we choose the \texttt{G4EmStandardPhysics\_option4} physics constructor, which is the most accurate one according to the \geant{} manual, and leave the production cut at its default setting of \qty{1}{\mm}. To prevent a systematic bias by the applied goodness-of-fit test on the results, we used three different tests (Anderson-Darling, Kolmogorov-Smirnov, and $\chi^2$); inconsistencies between their results justified this precaution.

We measure the performance of a given physics configuration by their efficiency, defined as the fraction of test cases for which the simulated spectra are compatible with the spectra from the reference physics configuration. The \texttt{G4EmStandardPhysics\_option1} and \texttt{G4EmStandardPhysics\_option2} physics constructors exhibited the worst compatibility with the reference simulations, while the \texttt{G4EmLivermore} physics constructor achieved the highest efficiency across all compatibility tests. Analysis based on $\chi^2$ tests imply that the efficiency of \texttt{G4EmLivermore} may be independent of the applied production cut. All studied physics constructors performed equally well for \cawo{} and \ce{Ge} targets, but generally performed worse in thin targets. As the energy deposition in thin targets is more sensitive to the details of the physics process and model implemented by the physics constructor, it seems plausible that for these cases the differences between the different physics constructors are more obvious. We found that in some cases a decreased production cut value improves the efficiency, but not in all cases.

Comparing different production cuts for a given physics constructor, we found that only for \texttt{G4EmStandard\-Physics\_option1} the production cut has unambiguously a significant impact on the efficiency. Marginalising over the production cut and comparing the total efficiencies of physics constructors, we found a set consisting of \texttt{G4EmPenelope}, \texttt{G4EmWenzelWVI}, \texttt{G4EmStandardPhysics\_option4}, and \texttt{G4EmLivermore} that shows no significant difference. However, \texttt{G4EmWenzelWVI} performs up to five times slower than the other three set members.

In a future work, it will be sufficient to validate, i.e.\ compare with experimental data, only one physics configuration out of the set of compatible ones to estimate the accuracy of all compatible physics configurations.

\backmatter

\bmhead{Acknowledgment}
This work has been funded through the Sonderforschungsbereich (Collaborative Research Center) SFB1258 ``Neutrinos and Dark Matter in Astro- and Particle Physics'', by Austria's Agency for Education and Internationalisation (OeAD) project SK 06/2018, and by the Austrian science fund (FWF): projects \href{http://dx.doi.org/10.55776/I5420}{DOI: 10.55776/I5420} and \href{http://dx.doi.org/10.55776/P34778}{DOI:10.55776/P34778}. The Bratislava group acknowledges partial support provided by the Slovak Research and Development Agency (project APVV-15-0576 and APVV-21-0377). This work was supported by the Ministry of Education, Youth and Sports of the Czech Republic under the Contract Number LM2023063. 
The computational results presented were partially obtained using the CLIP cluster and the Max Planck Computing and Data Facility (MPCDF). A. Fuß, H. Kluck, and V. Mokina gratefully recognise the support of J. Schieck, director of the Institut f\"ur Hochenergiephysik der \"Osterreichischen Akademie der Wissenschaften.

\bmhead{Declarations}
\begin{itemize}

\item \textbf{Conflict of interest/Competing interests:} The authors have no relevant financial or non-financial interests to disclose.
\item \textbf{Data availability:} Data sets generated during the current study are available from the corresponding author on reasonable request.
\item \textbf{Code availability:} The code used to generate the current study is available from the corresponding author on reasonable request.
\item \textbf{Author contribution:} The project on which this article is based was conceived by H. Kluck; he and R. Breier acquired the funding for it. H. Kluck developed the \geant{} simulation code and together with R. Breier, A. Fuß, and V. Mokina executed the simulations. V. Palušov\'a researched, implemented, and conducted the statistical analysis of the simulated data. V. Mokina researched the electromagnetic physics constructors of \geant{}; R. Breier studied their impact on the computing performances. H. Kluck, V. Mokina, V. Palušov\'a, R. Breier wrote the manuscript and created the figures; all authors reviewed and agreed to the article.
\end{itemize}

\begin{appendices}
\section{\label{appendix}Efficiencies of \geant{} configurations and their uncertainties}
\begin{figure*}
    \centering
    \includegraphics[height=0.92\textheight,keepaspectratio]{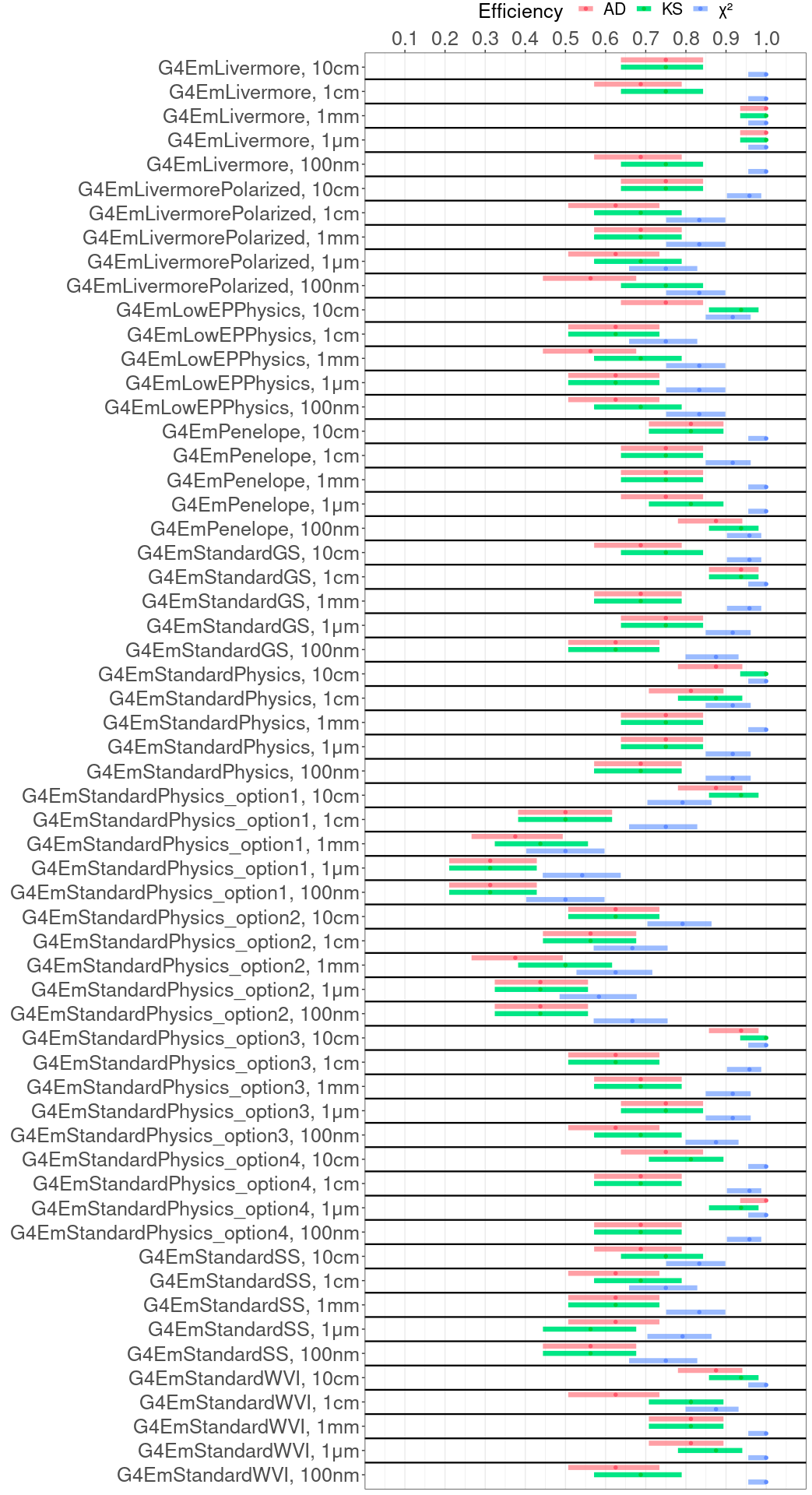}
    \caption{Efficiencies of \geant{} physics configurations, i.e.\ pairs of physics constructor and production cut value, and their uncertainties. The efficiencies are shown for
different goodness-of-fit tests: Anderson-Darling (AD), Kolmogorov-Smirnov (KS), and $\chi^2$.}
    \label{fig:uncertainties}
\end{figure*}

\Cref{fig:uncertainties} shows the efficiencies of all studied \geant{} physics configurations and their uncertainties, evaluated for all three goodness-of-fit tests. For details see \cref{sec:results:gof:total}.

\end{appendices}
\newpage
\bibliography{main} 


\begin{thebibliography}{48}
\ifx \bisbn   \undefined \def \bisbn  #1{ISBN #1}\fi
\ifx \binits  \undefined \def \binits#1{#1}\fi
\ifx \bauthor  \undefined \def \bauthor#1{#1}\fi
\ifx \batitle  \undefined \def \batitle#1{#1}\fi
\ifx \bjtitle  \undefined \def \bjtitle#1{#1}\fi
\ifx \bvolume  \undefined \def \bvolume#1{\textbf{#1}}\fi
\ifx \byear  \undefined \def \byear#1{#1}\fi
\ifx \bissue  \undefined \def \bissue#1{#1}\fi
\ifx \bfpage  \undefined \def \bfpage#1{#1}\fi
\ifx \blpage  \undefined \def \blpage #1{#1}\fi
\ifx \burl  \undefined \def \burl#1{\textsf{#1}}\fi
\ifx \doiurl  \undefined \def \doiurl#1{\url{https://doi.org/#1}}\fi
\ifx \betal  \undefined \def \betal{\textit{et al.}}\fi
\ifx \binstitute  \undefined \def \binstitute#1{#1}\fi
\ifx \binstitutionaled  \undefined \def \binstitutionaled#1{#1}\fi
\ifx \bctitle  \undefined \def \bctitle#1{#1}\fi
\ifx \beditor  \undefined \def \beditor#1{#1}\fi
\ifx \bpublisher  \undefined \def \bpublisher#1{#1}\fi
\ifx \bbtitle  \undefined \def \bbtitle#1{#1}\fi
\ifx \bedition  \undefined \def \bedition#1{#1}\fi
\ifx \bseriesno  \undefined \def \bseriesno#1{#1}\fi
\ifx \blocation  \undefined \def \blocation#1{#1}\fi
\ifx \bsertitle  \undefined \def \bsertitle#1{#1}\fi
\ifx \bsnm \undefined \def \bsnm#1{#1}\fi
\ifx \bsuffix \undefined \def \bsuffix#1{#1}\fi
\ifx \bparticle \undefined \def \bparticle#1{#1}\fi
\ifx \barticle \undefined \def \barticle#1{#1}\fi
\bibcommenthead
\ifx \bconfdate \undefined \def \bconfdate #1{#1}\fi
\ifx \botherref \undefined \def \botherref #1{#1}\fi
\ifx \url \undefined \def \url#1{\textsf{#1}}\fi
\ifx \bchapter \undefined \def \bchapter#1{#1}\fi
\ifx \bbook \undefined \def \bbook#1{#1}\fi
\ifx \bcomment \undefined \def \bcomment#1{#1}\fi
\ifx \oauthor \undefined \def \oauthor#1{#1}\fi
\ifx \citeauthoryear \undefined \def \citeauthoryear#1{#1}\fi
\ifx \endbibitem  \undefined \def \endbibitem {}\fi
\ifx \bconflocation  \undefined \def \bconflocation#1{#1}\fi
\ifx \arxivurl  \undefined \def \arxivurl#1{\textsf{#1}}\fi
\csname PreBibitemsHook\endcsname

\bibitem[\protect\citeauthoryear{Abdelhameed et~al.}{2019}]{CRESST:2019jnq}
\begin{barticle}
\bauthor{\bsnm{Abdelhameed}, \binits{A.H.}}, \betal:
\batitle{{First results from the {CRESST-III} low-mass dark matter program}}.
\bjtitle{Phys. Rev. D}
\bvolume{100}(\bissue{10}),
\bfpage{102002}
(\byear{2019})
\doiurl{10.1103/PhysRevD.100.102002}
{\href{https://arxiv.org/abs/1904.00498}{{arXiv:1904.00498}}}
{[astro-ph.CO]}
\end{barticle}
\endbibitem

\bibitem[\protect\citeauthoryear{Alkhatib et~al.}{2021}]{Alkhatib2021}
\begin{barticle}
\bauthor{\bsnm{Alkhatib}, \binits{I.}}, \betal:
\batitle{Constraints on lightly ionizing particles from {CDMSlite}}.
\bjtitle{Phys. Rev. Lett.}
\bvolume{127}(\bissue{8}),
\bfpage{081802}
(\byear{2021})
\doiurl{10.1103/PhysRevLett.127.081802}
{\href{https://arxiv.org/abs/2011.09183}{{arXiv:2011.09183}}}
{[hep-ex]}
\end{barticle}
\endbibitem

\bibitem[\protect\citeauthoryear{Angloher et~al.}{2019}]{Angloher2019}
\begin{barticle}
\bauthor{\bsnm{Angloher}, \binits{G.}}, \betal:
\batitle{Exploring {CE$\nu$NS} with {NUCLEUS} at the {Chooz} nuclear power plant}.
\bjtitle{Eur. Phys. J. C}
\bvolume{79}(\bissue{12}),
\bfpage{1018}
(\byear{2019})
\doiurl{10.1140/epjc/s10052-019-7454-4}
{\href{https://arxiv.org/abs/1905.10258}{{arXiv:1905.10258}}}
{[physics.ins-det]}
\end{barticle}
\endbibitem

\bibitem[\protect\citeauthoryear{Ackermann et~al.}{2024}]{Ackermann2024}
\begin{barticle}
\bauthor{\bsnm{Ackermann}, \binits{N.}}, \betal:
\batitle{Final {CONUS} results on coherent elastic neutrino-nucleus scattering at the {Brokdorf} reactor}.
\bjtitle{Phys. Rev. Lett.}
\bvolume{133},
\bfpage{251802}
(\byear{2024})
\doiurl{10.1103/PhysRevLett.133.251802}
{\href{https://arxiv.org/abs/2401.07684}{{arXiv:2401.07684}}}
{[hep-ex]}
\end{barticle}
\endbibitem

\bibitem[\protect\citeauthoryear{Abgrall et~al.}{2017}]{Abgrall2017}
\begin{barticle}
\bauthor{\bsnm{Abgrall}, \binits{N.}}, \betal:
\batitle{The large enriched germanium experiment for neutrinoless double beta decay ({LEGEND})}.
\bjtitle{AIP Conf. Proc.}
\bvolume{1894}(\bissue{1}),
\bfpage{020027}
(\byear{2017})
\doiurl{10.1063/1.5007652}
{\href{https://arxiv.org/abs/1709.01980}{{arXiv:1709.01980}}}
{[physics.ins-det]}
\end{barticle}
\endbibitem

\bibitem[\protect\citeauthoryear{Agostinelli et~al.}{2003}]{GEANT4:2002zbu}
\begin{barticle}
\bauthor{\bsnm{Agostinelli}, \binits{S.}}, \betal:
\batitle{{{GEANT4}--a simulation toolkit}}.
\bjtitle{Nucl. Instrum. Meth. Res. Sect. A}
\bvolume{506},
\bfpage{250}--\blpage{303}
(\byear{2003})
\doiurl{10.1016/S0168-9002(03)01368-8}
\end{barticle}
\endbibitem

\bibitem[\protect\citeauthoryear{Allison et~al.}{2006}]{Allison:2006ve}
\begin{barticle}
\bauthor{\bsnm{Allison}, \binits{J.}}, \betal:
\batitle{{{Geant4} developments and applications}}.
\bjtitle{{IEEE} Trans. Nucl. Sci.}
\bvolume{53},
\bfpage{270}
(\byear{2006})
\doiurl{10.1109/TNS.2006.869826}
\end{barticle}
\endbibitem

\bibitem[\protect\citeauthoryear{Allison et~al.}{2016}]{Allison:2016lfl}
\begin{barticle}
\bauthor{\bsnm{Allison}, \binits{J.}}, \betal:
\batitle{{Recent developments in {Geant4}}}.
\bjtitle{Nucl. Instrum. Meth. Res. Sect. A}
\bvolume{835},
\bfpage{186}--\blpage{225}
(\byear{2016})
\doiurl{10.1016/j.nima.2016.06.125}
\end{barticle}
\endbibitem

\bibitem[\protect\citeauthoryear{{Geant4 Collaboration}}{}]{Geant4Manual}
\begin{botherref}
\oauthor{\bsnm{{Geant4 Collaboration}}}:
Book For Application Developers: Release 10.6.
Rev4.1 - August 11th, 2020.
\url{https://geant4-userdoc.web.cern.ch/UsersGuides/ForApplicationDeveloper/BackupVersions/V10.6c/fo/BookForApplicationDevelopers.pdf}
\end{botherref}
\endbibitem

\bibitem[\protect\citeauthoryear{Guatelli et~al.}{2007}]{Guatelli2007}
\begin{barticle}
\bauthor{\bsnm{Guatelli}, \binits{S.}}, \betal:
\batitle{Validation of {Geant4} atomic relaxation against the nist physical reference data}.
\bjtitle{{IEEE} Trans. Nucl. Sci.}
\bvolume{54}(\bissue{3}),
\bfpage{594}--\blpage{603}
(\byear{2007})
\doiurl{10.1109/TNS.2007.894814}
\end{barticle}
\endbibitem

\bibitem[\protect\citeauthoryear{Lechner et~al.}{2009}]{Lechner2009}
\begin{barticle}
\bauthor{\bsnm{Lechner}, \binits{A.}},
\bauthor{\bsnm{Pia}, \binits{M.G.}},
\bauthor{\bsnm{Sudhakar}, \binits{M.}}:
\batitle{Validation of {Geant4} low energy electromagnetic processes against precision measurements of electron energy deposition}.
\bjtitle{{IEEE} Trans. Nucl. Sci.}
\bvolume{56}(\bissue{2}),
\bfpage{398}--\blpage{416}
(\byear{2009})
\doiurl{10.1109/tns.2009.2013858}
\end{barticle}
\endbibitem

\bibitem[\protect\citeauthoryear{Seo et~al.}{2011}]{Seo2011}
\begin{barticle}
\bauthor{\bsnm{Seo}, \binits{H.}}, \betal:
\batitle{Ionization cross sections for low energy electron transport}.
\bjtitle{{IEEE} Trans. Nucl. Sci.}
\bvolume{58}(\bissue{6}),
\bfpage{3219}--\blpage{3245}
(\byear{2011})
\doiurl{10.1109/tns.2011.2171992}
{\href{https://arxiv.org/abs/1110.2357}{{arXiv:1110.2357}}}
{[physics.comp-ph]}
\end{barticle}
\endbibitem

\bibitem[\protect\citeauthoryear{Batic et~al.}{2013}]{Batic2013}
\begin{barticle}
\bauthor{\bsnm{Batic}, \binits{M.}}, \betal:
\batitle{Validation of {Geant4} simulation of electron energy deposition}.
\bjtitle{{IEEE} Trans. Nucl. Sci.}
\bvolume{60}(\bissue{4}),
\bfpage{2934}--\blpage{2957}
(\byear{2013})
\doiurl{10.1109/tns.2013.2272404}
{\href{https://arxiv.org/abs/1307.0933}{{arXiv:1307.0933}}}
{[physics.comp-ph]}
\end{barticle}
\endbibitem

\bibitem[\protect\citeauthoryear{Basaglia et~al.}{2015}]{Basaglia2015}
\begin{barticle}
\bauthor{\bsnm{Basaglia}, \binits{T.}}, \betal:
\batitle{Investigation of {Geant4} simulation of electron backscattering}.
\bjtitle{{IEEE} Trans. Nucl. Sci.}
\bvolume{62}(\bissue{4}),
\bfpage{1805}--\blpage{1812}
(\byear{2015})
\doiurl{10.1109/tns.2015.2442292}
{\href{https://arxiv.org/abs/1506.01531}{{arXiv:1506.01531}}}
{[physics.comp-ph]}
\end{barticle}
\endbibitem

\bibitem[\protect\citeauthoryear{Basaglia et~al.}{2016}]{Basaglia2016a}
\begin{barticle}
\bauthor{\bsnm{Basaglia}, \binits{T.}}, \betal:
\batitle{Quantitative test of the evolution of {Geant4} electron backscattering simulation}.
\bjtitle{{IEEE} Trans. Nucl. Sci.}
\bvolume{63}(\bissue{6}),
\bfpage{2849}--\blpage{2865}
(\byear{2016})
\doiurl{10.1109/tns.2016.2617834}
{\href{https://arxiv.org/abs/1610.06349}{{arXiv:1610.06349}}}
{[physics.comp-ph]}
\end{barticle}
\endbibitem

\bibitem[\protect\citeauthoryear{Cirrone et~al.}{2010}]{Cirrone2010}
\begin{barticle}
\bauthor{\bsnm{Cirrone}, \binits{G.A.P.}}, \betal:
\batitle{Validation of the {Geant4} electromagnetic photon cross-sections for elements and compounds}.
\bjtitle{Nucl. Instrum. Methods Phys. Res. Sect. A}
\bvolume{618}(\bissue{1-3}),
\bfpage{315}--\blpage{322}
(\byear{2010})
\doiurl{10.1016/j.nima.2010.02.112}
\end{barticle}
\endbibitem

\bibitem[\protect\citeauthoryear{Arce et~al.}{2020}]{Arce:2020}
\begin{barticle}
\bauthor{\bsnm{Arce}, \binits{P.}}, \betal:
\batitle{Report on g4‐med, a geant4 benchmarking system for medical physics applications developed by the geant4 medical simulation benchmarking group}.
\bjtitle{Med. Phys.}
\bvolume{48}(\bissue{1}),
\bfpage{19}--\blpage{56}
(\byear{2020})
\doiurl{10.1002/mp.14226}
\end{barticle}
\endbibitem

\bibitem[\protect\citeauthoryear{Basaglia et~al.}{2015}]{Basaglia2015a}
\begin{barticle}
\bauthor{\bsnm{Basaglia}, \binits{T.}}, \betal:
\batitle{Experimental quantification of {Geant4} {PhysicsList} recommendations: methods and results}.
\bjtitle{J. Phys.: Conf. Ser.}
\bvolume{664}(\bissue{7}),
\bfpage{072037}
(\byear{2015})
\doiurl{10.1088/1742-6596/664/7/072037}
\end{barticle}
\endbibitem

\bibitem[\protect\citeauthoryear{Breier et~al.}{2023}]{IDM2022}
\begin{botherref}
\oauthor{\bsnm{Breier}, \binits{R.}}, et al.:
{Influence of Geant4 physics list on simulation accuracy and performance}.
SciPost Phys. Proc.,
063
(2023)
\doiurl{10.21468/SciPostPhysProc.12.063}
\end{botherref}
\endbibitem

\bibitem[\protect\citeauthoryear{{IAEA - Nuclear Data Section}}{}]{LiveChartOfNucllide}
\begin{botherref}
\oauthor{\bsnm{{IAEA - Nuclear Data Section}}}:
Live Chart of Nuclides.
\url{https://www-nds.iaea.org/relnsd/vcharthtml/VChartHTML.html}.
Accessed: 2022-11-02
\end{botherref}
\endbibitem

\bibitem[\protect\citeauthoryear{Ivanchenko et~al.}{2010}]{Ivanchenko2010}
\begin{barticle}
\bauthor{\bsnm{Ivanchenko}, \binits{V.N.}}, \betal:
\batitle{Geant4 models for simulation of multiple scattering}.
\bjtitle{J. Phys.: Conf. Ser.}
\bvolume{219}(\bissue{3}),
\bfpage{032045}
(\byear{2010})
\doiurl{10.1088/1742-6596/219/3/032045}
\end{barticle}
\endbibitem

\bibitem[\protect\citeauthoryear{Fernández-Varea et~al.}{1993}]{FernandezVarea1993}
\begin{botherref}
\oauthor{\bsnm{Fernández-Varea}, \binits{J.M.}}, et al.:
On the theory and simulation of multiple elastic scattering of electrons
(4),
447--473
(1993)
\doiurl{10.1016/0168-583x(93)95827-r}
\end{botherref}
\endbibitem

\bibitem[\protect\citeauthoryear{Wentzel}{1926}]{Wentzel1926}
\begin{botherref}
\oauthor{\bsnm{Wentzel}, \binits{G.}}:
{Zwei Bemerkungen über die Zerstreuung korpuskularer Strahlen als Beugungserscheinung}
\textbf{40}(8),
590--593
(1926)
\doiurl{10.1007/bf01390457}
\end{botherref}
\endbibitem

\bibitem[\protect\citeauthoryear{Bethe}{1953}]{Bethe1953}
\begin{botherref}
\oauthor{\bsnm{Bethe}, \binits{H.A.}}:
Molière’s theory of multiple scattering
\textbf{89}(6),
1256--1266
(1953)
\doiurl{10.1103/physrev.89.1256}
\end{botherref}
\endbibitem

\bibitem[\protect\citeauthoryear{Boschini et~al.}{2013}]{Boschini2013}
\begin{barticle}
\bauthor{\bsnm{Boschini}, \binits{M.J.}}, \betal:
\batitle{An expression for the mott cross section of electrons and positrons on nuclei with z up to 118}.
\bjtitle{Radiat. Phys. Chem.}
\bvolume{90},
\bfpage{39}--\blpage{66}
(\byear{2013})
\doiurl{10.1016/j.radphyschem.2013.04.020}
{\href{https://arxiv.org/abs/1304.5871}{{arXiv:1304.5871}}}
{[physics.atom-ph]}
\end{barticle}
\endbibitem

\bibitem[\protect\citeauthoryear{Urbán}{}]{Urban2006}
\begin{botherref}
\oauthor{\bsnm{Urbán}, \binits{L.}}:
A model for multipel scattering in geant4.
Technical Report CERN-OPEN-2006-077,
CERN.
\url{https://cds.cern.ch/record/1004190/files/open-2006-077.pdf}
\end{botherref}
\endbibitem

\bibitem[\protect\citeauthoryear{Lewis}{}]{Lewis1950}
\begin{botherref}
\oauthor{\bsnm{Lewis}, \binits{H.W.}}:
Multiple scattering in an infinite medium
\textbf{78}(5),
526--529
\doiurl{10.1103/physrev.78.526}
\end{botherref}
\endbibitem

\bibitem[\protect\citeauthoryear{Novak}{2025}]{Novak:2025}
\begin{botherref}
\oauthor{\bsnm{Novak}, \binits{M.}}:
On the new and accurate (Goudsmit-Saunderson) model for describing e-/e+ multiple Coulomb scattering (Geant4 Technical Note)
(2025).
\url{https://arxiv.org/abs/2410.13361}
\end{botherref}
\endbibitem

\bibitem[\protect\citeauthoryear{{Geant4 Collaboration}}{}]{Geant4PhysicsManual}
\begin{botherref}
\oauthor{\bsnm{{Geant4 Collaboration}}}:
Physics Reference Manual: Release 10.6.
Rev4.0: December 6th, 2019.
\url{https://geant4-userdoc.web.cern.ch/UsersGuides/PhysicsReferenceManual/BackupVersions/V10.6/fo/PhysicsReferenceManual.pdf}
\end{botherref}
\endbibitem

\bibitem[\protect\citeauthoryear{Apostolakis et~al.}{2009}]{Apostolakis:2009egq}
\begin{barticle}
\bauthor{\bsnm{Apostolakis}, \binits{J.}}, \betal:
\batitle{{Geometry and physics of the Geant4 toolkit for high and medium energy applications}}.
\bjtitle{Radiat. Phys. Chem.}
\bvolume{78}(\bissue{10}),
\bfpage{859}--\blpage{873}
(\byear{2009})
\doiurl{10.1016/j.radphyschem.2009.04.026}
\end{barticle}
\endbibitem

\bibitem[\protect\citeauthoryear{Dotti et~al.}{2011}]{Dotti:2011zz}
\begin{barticle}
\bauthor{\bsnm{Dotti}, \binits{A.}}, \betal:
\batitle{{Recent improvements on the description of hadronic interactions in Geant4}}.
\bjtitle{J. Phys. Conf. Ser.}
\bvolume{293},
\bfpage{012022}
(\byear{2011})
\doiurl{10.1088/1742-6596/293/1/012022}
\end{barticle}
\endbibitem

\bibitem[\protect\citeauthoryear{Kim}{2015}]{kimjae2015}
\begin{botherref}
\oauthor{\bsnm{Kim}, \binits{J.}}:
{How to Choose the Level of Significance: A Pedagogical Note}.
MPRA Paper 66373,
University Library of Munich, Germany
(2015).
\url{https://ideas.repec.org/p/pra/mprapa/66373.html}
\end{botherref}
\endbibitem

\bibitem[\protect\citeauthoryear{Kolmogorov}{1933}]{Kolmogorov1933}
\begin{barticle}
\bauthor{\bsnm{Kolmogorov}, \binits{A.L.}}:
\batitle{Sulla determinazione empirica di una legge di distribuzione}.
\bjtitle{Giornale dell'Istituto Italiano Degli Attuari}
\bvolume{4},
\bfpage{83}--\blpage{91}
(\byear{1933})
\end{barticle}
\endbibitem

\bibitem[\protect\citeauthoryear{Smirnov}{1939}]{Smirnov1939}
\begin{barticle}
\bauthor{\bsnm{Smirnov}, \binits{N.V.}}:
\batitle{On the estimation of discrepancy between empirical curves of distribution for two independent samples}.
\bjtitle{Mosc. Math. Bull.}
\bvolume{2},
\bfpage{3}--\blpage{14}
(\byear{1939})
\end{barticle}
\endbibitem

\bibitem[\protect\citeauthoryear{Pearson}{1900}]{Pearson1900}
\begin{barticle}
\bauthor{\bsnm{Pearson}, \binits{K.}}:
\batitle{X. on the criterion that a given system of deviations from the probable in the case of a correlated system of variables is such that it can be reasonably supposed to have arisen from random sampling}.
\bjtitle{Lond. Edinb. Dubl. Phil. Mag.}
\bvolume{50}(\bissue{302}),
\bfpage{157}--\blpage{175}
(\byear{1900})
\doiurl{10.1080/14786440009463897}
\end{barticle}
\endbibitem

\bibitem[\protect\citeauthoryear{Cochran}{1952}]{Cochran1952}
\begin{barticle}
\bauthor{\bsnm{Cochran}, \binits{W.G.}}:
\batitle{{The $\chi^2$ Test of Goodness of Fit}}.
\bjtitle{Ann. Math. Stat.}
\bvolume{23}(\bissue{3}),
\bfpage{315}--\blpage{345}
(\byear{1952})
\doiurl{10.1214/aoms/1177729380}
\end{barticle}
\endbibitem

\bibitem[\protect\citeauthoryear{Anderson and Darling}{1954}]{Anderson1956}
\begin{barticle}
\bauthor{\bsnm{Anderson}, \binits{T.W.}},
\bauthor{\bsnm{Darling}, \binits{D.A.}}:
\batitle{A test of goodness of fit}.
\bjtitle{J. Am. Stat. Assoc.}
\bvolume{49}(\bissue{268}),
\bfpage{765}--\blpage{769}
(\byear{1954})
\doiurl{10.1080/01621459.1954.10501232}
\end{barticle}
\endbibitem

\bibitem[\protect\citeauthoryear{Kim et~al.}{2015}]{Kim2015}
\begin{barticle}
\bauthor{\bsnm{Kim}, \binits{S.H.}}, \betal:
\batitle{Validation test of {Geant4} simulation of electron backscattering}.
\bjtitle{{IEEE} Trans. Nucl. Sci.}
\bvolume{62}(\bissue{2}),
\bfpage{451}--\blpage{479}
(\byear{2015})
\doiurl{10.1109/tns.2015.2401055}
{\href{https://arxiv.org/abs/1502.01507}{{arXiv:1502.01507}}}
{[physics.comp-ph]}
\end{barticle}
\endbibitem

\bibitem[\protect\citeauthoryear{Paterno}{2004}]{Paterno:2004cb}
\begin{botherref}
\oauthor{\bsnm{Paterno}, \binits{M.}}:
Calculating efficiencies and their uncertainties
(2004)
\doiurl{10.2172/15017262} .
FERMILAB-TM-2286-CD
\end{botherref}
\endbibitem

\bibitem[\protect\citeauthoryear{Fisher}{1922}]{fisher}
\begin{barticle}
\bauthor{\bsnm{Fisher}, \binits{R.A.}}:
\batitle{On the interpretation of $\chi^2$ from contingency tables, and the calculation of p}.
\bjtitle{J. R. Stat. Soc.}
\bvolume{85}(\bissue{1}),
\bfpage{87}--\blpage{94}
(\byear{1922})
\doiurl{10.2307/2340521}
\end{barticle}
\endbibitem

\bibitem[\protect\citeauthoryear{McNemar}{1947}]{McNemar}
\begin{botherref}
\oauthor{\bsnm{McNemar}, \binits{Q.}}:
{Note on the sampling error of the difference between correlated proportions or percentages}.
Psychometrika
\textbf{12}
(1947)
\doiurl{10.1007/BF02295996}
\end{botherref}
\endbibitem

\bibitem[\protect\citeauthoryear{Yates}{1934}]{Yates1934}
\begin{barticle}
\bauthor{\bsnm{Yates}, \binits{F.}}:
\batitle{Contingency tables involving small numbers and the χ2 test}.
\bjtitle{Suppl. J. R. Stat. Soc.}
\bvolume{1}(\bissue{2}),
\bfpage{217}--\blpage{235}
(\byear{1934})
\doiurl{10.2307/2983604}
\end{barticle}
\endbibitem

\bibitem[\protect\citeauthoryear{Brun and Rademakers}{1997}]{Brun1997}
\begin{barticle}
\bauthor{\bsnm{Brun}, \binits{R.}},
\bauthor{\bsnm{Rademakers}, \binits{F.}}:
\batitle{{{ROOT}: An object oriented data analysis framework}}.
\bjtitle{Nucl. Instrum. Meth. Res. Sect. A}
\bvolume{389},
\bfpage{81}--\blpage{86}
(\byear{1997})
\doiurl{10.1016/S0168-9002(97)00048-X}
\end{barticle}
\endbibitem

\bibitem[\protect\citeauthoryear{{R Core Team}}{2022}]{R}
\begin{bbook}
\bauthor{\bsnm{{R Core Team}}}:
\bbtitle{R: A Language and Environment for Statistical Computing}.
\bpublisher{R Foundation for Statistical Computing},
\blocation{Vienna, Austria}
(\byear{2022}).
\bcomment{R Foundation for Statistical Computing}.
\burl{https://www.R-project.org/}
\end{bbook}
\endbibitem

\bibitem[\protect\citeauthoryear{Muškinja et~al.}{2020}]{Muskinja2020}
\begin{barticle}
\bauthor{\bsnm{Muškinja}, \binits{M.}},
\bauthor{\bsnm{Chapman}, \binits{J.D.}},
\bauthor{\bsnm{Gray}, \binits{H.}}:
\batitle{Geant4 performance optimization in the {ATLAS} experiment}.
\bjtitle{EPJ Web Conf.}
\bvolume{245},
\bfpage{02036}
(\byear{2020})
\doiurl{10.1051/epjconf/202024502036}
\end{barticle}
\endbibitem

\bibitem[\protect\citeauthoryear{Dotti et~al.}{2015}]{Dotti2015}
\begin{barticle}
\bauthor{\bsnm{Dotti}, \binits{A.}}, \betal:
\batitle{Geant4 computing performance benchmarking and monitoring}.
\bjtitle{J. Phys.: Conf. Ser.}
\bvolume{664}(\bissue{6}),
\bfpage{062021}
(\byear{2015})
\doiurl{10.1088/1742-6596/664/6/062021}
\end{barticle}
\endbibitem

\bibitem[\protect\citeauthoryear{{Max-Planck-Gesellschaft}}{}]{MPCDF}
\begin{botherref}
\oauthor{\bsnm{{Max-Planck-Gesellschaft}}}:
The Max Planck Computing and Data Facility.
\url{https://www.mpcdf.mpg.de/}.
Accessed: 2023-05-25
\end{botherref}
\endbibitem

\bibitem[\protect\citeauthoryear{{Cloud Infrastructure Project}}{}]{CLIP}
\begin{botherref}
\oauthor{\bsnm{{Cloud Infrastructure Project}}}:
The Cloud Infrastructure Platform.
\url{https://www.clip.science/}.
Accessed: 2023-05-25
\end{botherref}
\endbibitem

\end{thebibliography}

\end{document}